\documentclass[11pt,sort&compress]{elsarticle}

\usepackage{amssymb}
\usepackage{amsmath}
\usepackage{graphicx}
\usepackage{tabularx}
\usepackage{cite}
\usepackage{color}
\usepackage{fullpage}
\usepackage{subfigure}
\usepackage{subfigmat}
\usepackage{etoolbox}
\usepackage{url}
\usepackage{float}
\usepackage{bm}
\usepackage[dvipsnames]{xcolor}
\usepackage{placeins}
\usepackage{siunitx} 
\usepackage[nameinlink,capitalize]{cleveref}
\newcommand{\new}[1]{{\color{black}#1}}
\newcommand{\newnew}[1]{{\color{black}#1}}
\journal{Wear}
\begin{document}

\begin{frontmatter}
  \title{Numerical prediction of erosion due to a cavitating jet}

  \author{Theresa Trummler$^{1,2}$}\ead{theresa.trummler@unibw.de}
  \author{Steffen J. Schmidt$^{1}$}
  \author{Nikolaus A. Adams$^{1}$}

  \address{
      $^{1}$Chair of Aerodynamics and Fluid Mechanics, Technical University of Munich \\ 
      Boltzmannstr.\ 15, 85748 Garching bei M\"unchen, Germany\\ 
      $^{2}$Institute of Applied Mathematics and Scientific Computing, Bundeswehr University Munich\\ Werner-Heisenberg-Weg 39, 85577 Neubiberg, Germany }

  \begin{abstract}

      We numerically investigate the erosion potential of a cavitating liquid jet by means of high-resolution finite volume simulations. As thermodynamic model, we employ a barotropic equilibrium cavitation approach, embedded into a homogeneous mixture model. To resolve the effects of collapsing vapor structures and to estimate the erosion potential\new{, full compressibility is considered.}

      Two different operating points featuring different cavitation intensities are investigated and their erosion potential is estimated and compared. Different methods are used for this purpose, including collapse detection \citep{Mihatsch:2015db}, maximum pressure distribution on the wall, and a new method of generating numerical pit equivalents. The data of numerical pit equivalents is analyzed in detail and compared with experimental data \new{from the literature}. Furthermore, a comprehensive grid study for both operating points is presented.

  \end{abstract}

  \begin{keyword}
     Cavitating jet; Numerical cavitation erosion prediction; Cavitation erosion; Cavitation pit; Grid study;
  \end{keyword}

\end{frontmatter}

\section{Introduction}\label{s:Introduction}

  Cavitation is the formation and subsequent collapse of cavities in fluids, both processes being initiated by pressure fluctuations in the surrounding fluid. When such collapses occur near a wall, shock waves emitted at collapse~\citep{rayleigh1917,hickling64} impact near surfaces and can cause material damage~\citep{benjamin66,Plesset:1971hu,Philipp:1998eg}. During the initial stage of cavitation erosion, referred to as the incubation period, the damage consists of circular, isolated plastic surface deformations called \textit{pits}~\citep{franc2014pitting}. Later stages of cavitation erosion then result in increased fracture and weight loss~\citep{chahine2014mass}. 

  Experimental pitting tests, first introduced by \citet{knapp1955recent, knapp1958accelerated}, have been established for assessing erosion aggressiveness during the incubation period. The idea behind pitting tests is to use the material itself as a kind of sensor directly recording erosive collapse events. The distribution of local pit-shaped surface deformations allows for visual and qualitative detection of erosion-prone areas. In addition, extensive quantitative evaluations are possible of e.g. pit distributions, pitting rates, cumulative rates over pit diameter or over pit depth, see \citet{franc2012material}. Pitting tests have also been standardized and are part of the American Society for Testing and Materials (ASTM) standards, such as the G-134 Test Method for Erosion of Solid Materials by a Cavitating Liquid Jet. Despite the importance of pitting tests in experimental studies, numerical approaches to obtain data comparable to these tests are missing. 

  In the last decade, numerical simulations have become a complementary approach to cavitation erosion prediction. With the help of computational fluid dynamics, detailed flow information can be obtained allowing to correlate flow dynamics and erosion mechanisms. Considering the compressibility of both phases, the pressure waves generated during collapse are resolved and thus are collapse-induced surface loads. \new{For example, compressible flow solvers have been used to study the pressure loads induced by near-wall bubble collapses, see e.g. \citep{Johnsen:2009cua,Lauer:2012jh}. Furthermore, \citet{Pohl:2015keb} and \citet{kaufhold2017numerical} conducted finite-element-assisted pit analysis using the hydrodynamic pressure data from compressible simulations.
  \citet{hsiao2014modelling} directly coupled a compressible finite difference flow solver with a boundary element method to study the stresses in the material and, more recently, e.g. \citet{sarkar2021fluid} also presented coupled simulations of near-wall bubble collapses. \citet{joshi2019sph,joshi2020bubble} employed a Smoothed Particle Hydrodynamics solver for coupled investigations.} For erosion prediction in complex configurations, a pioneering study was conducted by \citet{Mihatsch:2015db}. They numerically investigated the cavitation erosion potential in a cavitation loop using a fully compressible flow solver in combination with a homogeneous mixture model and an equilibrium phase change cavitation model~\citep{Schmidt:2015wa}. To obtain more detailed information about the collapse dynamics, they introduced a collapse detection algorithm that detects collapsing vapor structures and records the collapse pressure. For the evaluation of surface loads and validation with experimental data, \citet{Mihatsch:2015db} generated force load collectives - an evaluation method also recently taken up by \citet{schreiner20203d}. Using the same thermodynamic modeling approach as \citet{Mihatsch:2015db}, the erosion potential of cavitating flows has been studied in nozzles~\citep{Egerer:2014wu}, in throttle valves~\citep{Beban:2017vo}, in injector components~\citep{Koukouvinis:2016et, Orley:2016db}\new{,} on ship propellers~\citep{budich2015numerical}\new{, on hydrofoils~\citep{blume20193d}, and on ultrasonic horns~\citep{schreiner2021assessment,mottyll2016numerical}.} In recent years, research has also focused on incompressible approaches capable of predicting erosive aggressiveness as e.g. successfully achieved by \citet{Schenke:2019il, Arabnejad:2021bc} \new{and recently applied to URANS by \citet{melissaris2020accuracy}.}

  For numerical erosion prediction \new{in more complex configurations} predominantly qualitative methods are used. Mainly the distribution of an erosion-relevant quantity, or a thereof derived indicator, is considered. E.g. pressure peaks recorded on surfaces are often used, as for near-wall bubble collapses \citet{Lauer:2012jh,Trummler:2020JFM, Trummler:2021if, Pishchalnikov:2018pp}, in cavitating nozzle flows~\citet{Egerer:2014wu}, in injector components, see e.g.~\citep{Koukouvinis:2016et, Cristofaro:2019jr}, in throttle valves~\citep{Beban:2017vo} and on ship propellers~\citep{budich2015numerical}. However, without any further postprocessing, pressure peaks are of qualitative nature. Combining the maximum wall pressure distribution with detected collapse events~\citep{Mihatsch:2015db} leads to a more reliable prediction, see e.g.~\citet{Egerer:2014wu,Beban:2017vo}. Other quantities employed are for instance temporal derivatives of pressure and vapor~\citep{Koukouvinis:2015gt} or normalized impact power for incompressible simulations~\citep{Arabnejad:2021bc,Schenke:2019il}. Nevertheless, all these methods are of qualitative nature and quantitative comparison with experimental data from pitting tests can be challenging. To address this issue, we present a new approach for generating numerical pit equivalents (NPEs). Since grid resolution can strongly influence erosion-relevant quantities~\citep{schmidt2014assessment,Mihatsch:2015db,Trummler:2021if}, we also present a detailed grid study in this paper. 
  
  The considered configuration is a cavitating liquid jet, which belongs to the standard test methods for the erosion of solid materials (ASTM G134 - 17). Similar setups have been \new{experimentally} studied by e.g.~\citet{Berger:1983Erosion,kleinbreuer1979untersuchung,Watanabe:2015dk,Watanabe:2016dk,Fujisawa:2017jd,Fujisawa:2019ht}. We carry out fully compressible high-resolution simulations of two operating points covering different erosion intensities. Therefore, we employ an explicit, density-based method specifically designed for simulating cavitating flows~\citep{Schmidt:2015wa} that has been successfully validated for cavitation dynamics~\citep{Egerer:2014wu, budich2018jfm,Orley:2015kt} and numerical cavitation erosion prediction~\citep{Mihatsch:2015db,budich2015numerical} against experimental data. 

  The paper is structured as follows. In \cref{s:Methods} the governing equations, the thermodynamic models and the numerical approach are introduced. \Cref{s:Setup} presents the considered configuration and \Cref{s:NumSetup} the numerical setup. In \cref{s:Results}, we analyze the cavitation and collapse dynamics and assess the erosion potential using different methods. The grid dependence of the simulation results and cavitation erosion-relevant aspects is evaluated in \cref{s:Grid}. Finally, \cref{s:Conclusion} summarizes the paper. 

\section{Physical Model and Numerical Method}\label{s:Methods}

  \subsection{Governing Equations}\label{ss:GovEq}

    \new{We consider the fully compressible Navier-Stokes equations in conservative form 
        \begin{equation}
            \partial_{t}\boldsymbol{U}+\nabla \cdot [ \boldsymbol{C}(\boldsymbol{U})+\boldsymbol{S}(\boldsymbol{U})]=0\,, 
            \label{eq:NS}
        \end{equation}
    where $\boldsymbol{U}=[\rho , \, \rho \boldsymbol{u} ]^T$ is composed of the conserved variables density $\rho$ and momentum $\rho \boldsymbol{u}$. We model the liquid with a barotropic equation of state ($p=p(\rho)$) and therefore can omit the energy equation. $\boldsymbol{C}(\boldsymbol{U})$ denotes the convective fluxes $\boldsymbol{C}(\boldsymbol{U})$ and $\boldsymbol{S}(\boldsymbol{U})$ the flux contributions due to pressure and shear.  $\boldsymbol{C}(\boldsymbol{U})$ and $\boldsymbol{S}(\boldsymbol{U})$ read
    \begin{equation}
       \boldsymbol{C}(\boldsymbol{U})=
                \boldsymbol{u}
                   \begin{bmatrix} 
                      \rho\\ 
                      \rho\boldsymbol{u}\\ 
                \end{bmatrix}
                \quad
                \mathrm{and}
                \quad
                \boldsymbol{S}(\boldsymbol{U})=
                \begin{bmatrix} 
                0\\ 
                 p \boldsymbol{I}-\boldsymbol{\tau}\\ 
                \end{bmatrix},
                     \label{eq:NS_Basis}
        \end{equation}    
    where $\boldsymbol{u}$, $p$, $\boldsymbol{I}$, $\boldsymbol{\tau}$ denotes the velocity, the static pressure, the unit tensor, and the viscous stress tensor, respectively. $\boldsymbol{\tau}$ is calculated with 
    \begin{equation}
      \boldsymbol{\tau}=\mu(\nabla \boldsymbol{u}+(\nabla \boldsymbol{u})^{T}-\frac{2}{3}(\nabla \cdot\boldsymbol{u})\boldsymbol{I}),
      \label{eq:tau}
    \end{equation}
    where $\mu$ is the dynamic viscosity. }

  \subsection{Thermodynamic Model}\label{ss:ThermoModel}

    \new{Assuming that the liquid and gas phase of a cavitating liquid are in thermal and mechanical equilibrium, \citet{Schnerr:2008jja} formulated a cavitation model, which has been successfully validated for bubble collapses \citep{Egerer:2016it,Pohl:2015keb, Koukouvinis:2016ir}, }cavitating nozzle flows~\citep{Egerer:2014wu, Orley:2015kt,budich2018jfm} and numerical cavitation erosion prediction~\citep{Mihatsch:2015db,schmidt2014assessment,Trummler:2021if}. 

    Due to the thermodynamic equilibrium assumption for the cavitation model, finite-rate mass transfer terms are avoided. The liquid starts to cavitate if the pressure drops beneath saturation pressure $p_\mathrm{sat}$
         \begin{equation}
            p<p_\mathrm{sat}
            \label{eq:p}
        \end{equation}
    and then a liquid-vapor mixture is present. The cavitating liquid is described with an isentropic equation of state $p=p(\rho) \big|_{s=const.}$ stored in tabulated form. In the table also the vapor volume fraction $\alpha$, the speed of sound $c$ and the dynamic viscosity $\mu$ are included. Linear interpolation is used in between two tabulated values.

\begin{figure}[!tb]
      \centering
    \includegraphics[width=0.5\linewidth]{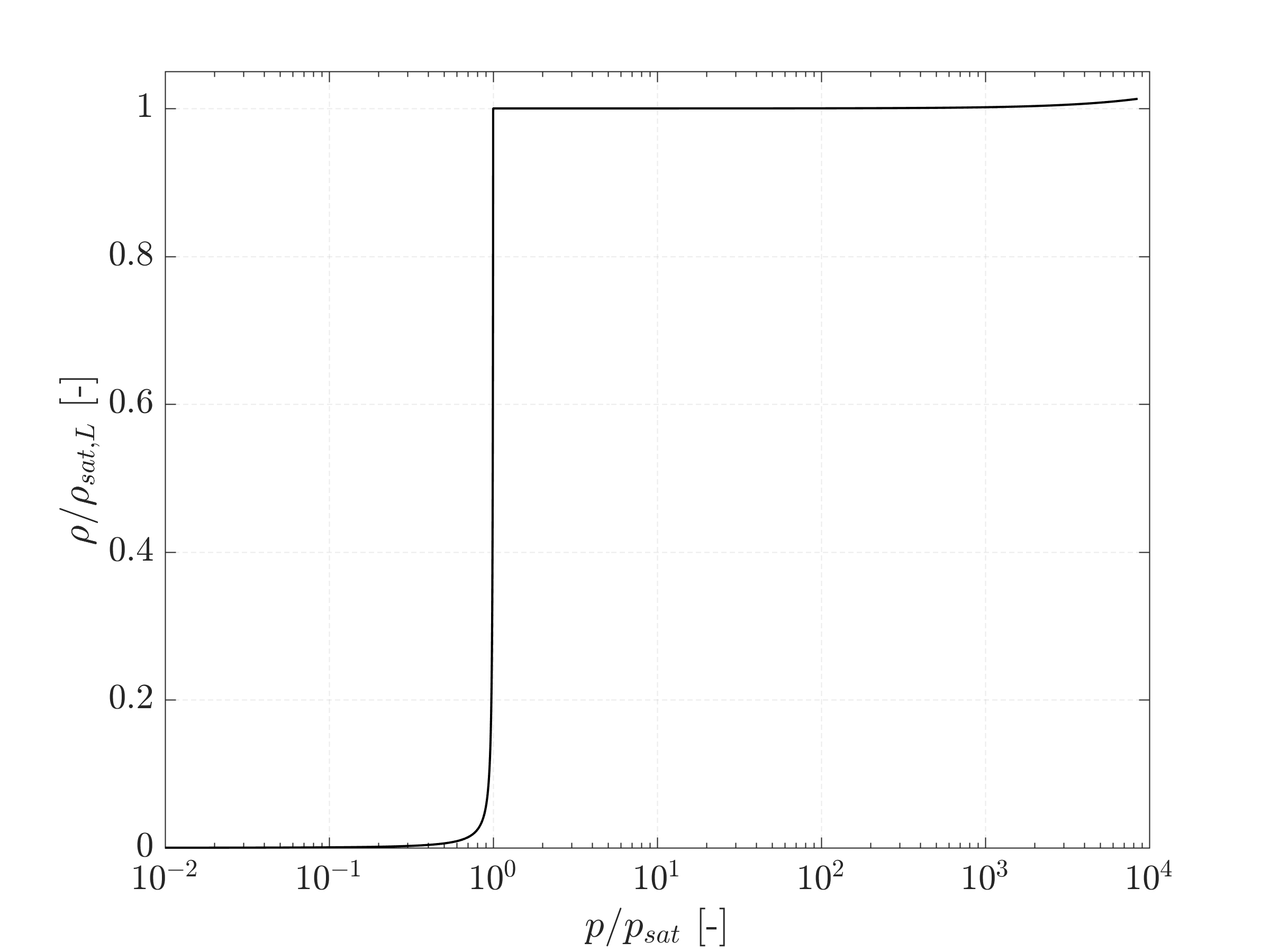}
      \caption{Isentropic relationship between pressure and density for the Diesel-like test fluid. }
      \label{fig:rho_p}
    \end{figure}

    As working fluid we use a Diesel-like test fluid. \Cref{fig:rho_p} shows the normalized density plotted over the normalized pressure. Characteristic properties of the test fluid at the reference temperature $T_{ref}$ are provided in Table~\ref{tab:eos}. The mixture viscosity was derived by linear blending with the volume fractions as 
        \begin{equation}
            \mu=\alpha\cdot\mu_{sat,V}+(1-\alpha)\cdot\mu_{sat,L},
            \label{eq:muLiqMix}
        \end{equation}
    where $\mu_{sat,V}$ , $\mu_{sat,L}$ denote the viscosity of the vapor and liquid, respectively. The used values are listed in Table~\ref{tab:eos}.
    Surface tension is neglected in our model as it has little qualitative relevance at the high pressures considered~\citep{Franc:2004fu}.

    \begin{table}[!tb]
      \centering
      \caption{Properties of the Diesel-like test fluid at $T_{ref}$.}
        \begin{tabular}{|cccccc|}
        \hline
         \textbf{$\boldsymbol{p_\mathrm{sat}}\textbf{\,[\si{Pa}]}$} & 
         \textbf{$\boldsymbol{\rho_{sat,L}}\textbf{\,[\si{kg/m^3}]}$} & 
         \textbf{$\boldsymbol{\rho_{sat,V}}\textbf{\,[\si{kg/m^3}]}$} & 
         \textbf{$\boldsymbol{c_L}\textbf{\,[\si{m/s}]}$} & 
         \textbf{$\boldsymbol{\mu_{sat,L}}\textbf{\,[\si{m Pa s}]}$} & 
         \textbf{$\boldsymbol{\mu_{sat,V}}\textbf{\,[\si{\mu Pa s}]}$}  \\
        \hline
        2371 & 815 & 0.0015 & 1000 & 2.3 & 19 \\
        \hline
        \end{tabular}
      \label{tab:eos}
    \end{table}

  \subsection{Numerical Method}\label{ss:NumMethod}

   \new{We employ a }density-based fully compressible flow solver with a low-Mach-number-consistent flux function, see \citet{Schmidt:2015wa}. \new{The reconstruction at the cell faces is realized with an upwind biased scheme, where the} velocity components are reconstructed with the up to third-order-accurate limiter of~\citet{Koren:1993} and the thermodynamic quantities $\rho$, $p$ with the second-order minmod slope limiter of~\citet{Roe:1986}. \new{An implicit modelling of the subgrid scale terms is achieved due to the nonlinearity of the shock capturing scheme for the convective fluxes, see also~\citet{garnier1999use,garnier2009large}. }Time integration is performed with an explicit second-order, 4-step low-storage Runge-Kutta method \citep{Schmidt:2015wa}. 

\section{Considered Configuration}\label{s:Setup} 

\begin{figure}[!htb]
    \centering
    \subfigure[]{\includegraphics[height=5cm]{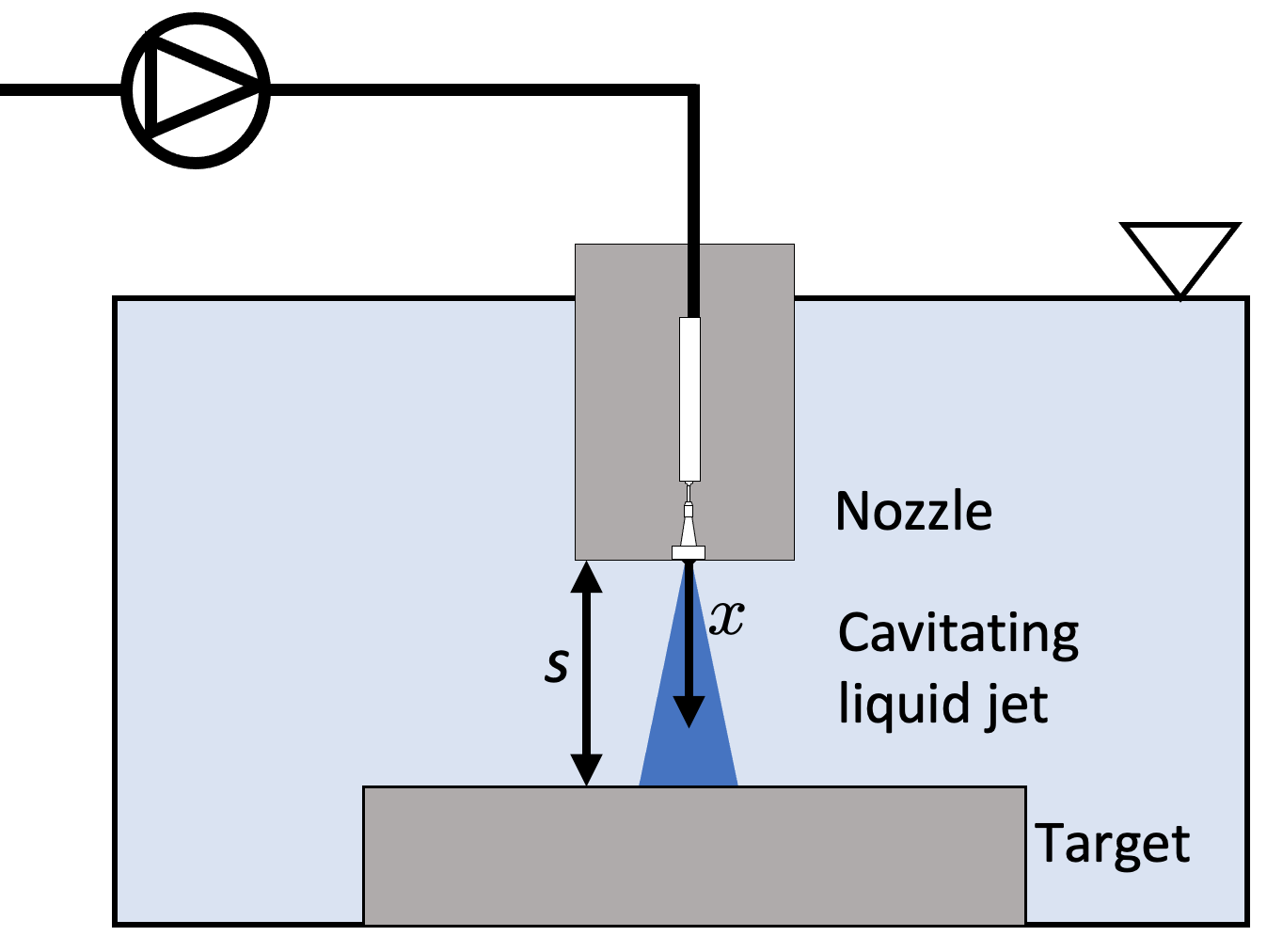}}
    \subfigure[]{\includegraphics[height=5cm]{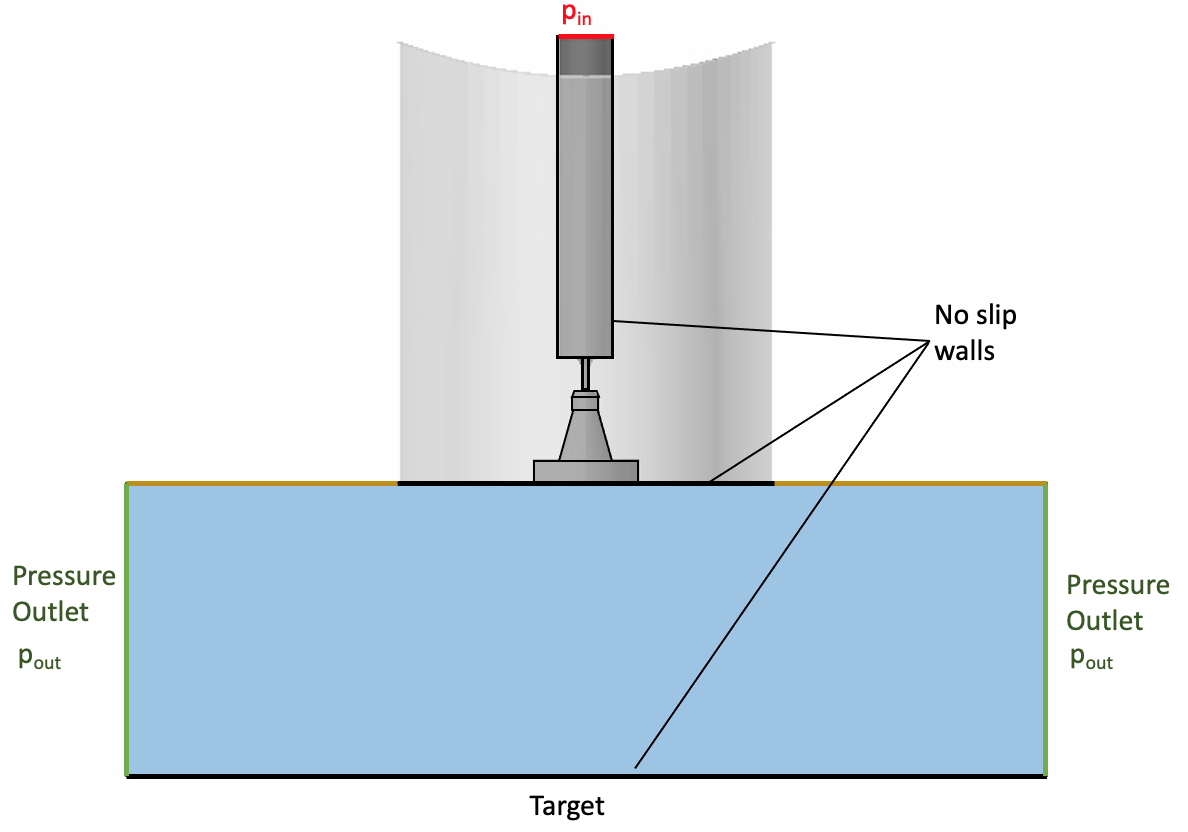}}
    \caption{Sketch of \new{a corresponding experiment of a cavitating jet} (a) and numerical setup (b).}
    \label{fig:setup}
  \end{figure}

  We consider a cavitating liquid jet, which is one of the standard test methods for the erosion of solid materials~(ASTM G134 - 17). Similar setups have been \new{experimentally} studied by e.g. \citet{Berger:1983Erosion,kleinbreuer1979untersuchung,Watanabe:2015dk,Watanabe:2016dk,Fujisawa:2017jd,Fujisawa:2019ht}. \Cref{fig:setup}~(a) shows a sketch \new{a corresponding experiment of a cavitating jet}. \new{The} test fluid is discharged through a nozzle into a liquid environment. Due to the high velocity of the jet, cavitation occurs and a cavitating jet forms. Near the target, the vapor clouds collapse and lead to ring-shaped material damage. \new{The formation of the cavity, its detachment and collapse is a periodic process with repetitive shedding cycles~\citep{Watanabe:2016dk,Fujisawa:2017jd,Fujisawa:2019ht}.}

  Different cavitation intensities are obtained by adjusting the inlet pressure $p_\mathrm{in}$ and thus the velocity of the jet. The intensity can be expressed by the cavitation number
  \begin{equation}  
       \sigma= \frac{p_\mathrm{out}-p_\mathrm{sat}}{p_\mathrm{in}-p_\mathrm{out}},
      \label{eq:f_general}
  \end{equation}
  where $p_\mathrm{in}$ denotes the inlet pressure, $p_\mathrm{out}$ the pressure in the chamber and $p_\mathrm{sat}$ vapor pressure of the liquid. The pressure difference $p_\mathrm{in}-p_\mathrm{out}$ determines the jet velocity and thus the cavitation intensity and $p_\mathrm{out}-p_\mathrm{sat}$ governs the lifetime of the vapor structures in the cavitating liquid jet. Therefore, as a function of $p_\mathrm{in}-p_\mathrm{out}$ and $p_\mathrm{out}-p_\mathrm{sat}$, there is a certain distance downstream the nozzle at which the implosions of the vapor structures are most energetic. At given pressure differences, the erosion aggressiveness varies with the distance between the target and the nozzle~\citep{Berger:1983Erosion,Watanabe:2016dk,Fujisawa:2017jd,Fujisawa:2019ht}, too. 

  \new{
  The turbulent Reynolds number represents a characteristic dimensionless number for jets and is determined with 
    \begin{equation}  
       Re_d= \frac{\rho\bar{u}d}{\mu},
      \label{eq:f_general}
  \end{equation}
  where $d$ denotes the throat diameter and $\bar{u}$ the mean velocity. For the considered cases, $Re_d$ is in the range of 30,000. 
  } 

  \begin{table}
    \caption{Overview of the operating points investigated.}
    \centering\begin{tabular}{|ccccc|}
    \hline
      \textbf{OP}& 
      $\boldsymbol{\sigma}$        \textbf{[-]} &  
      $\boldsymbol{p_\mathrm{in}}$\textbf{[bar]} & 
      $\boldsymbol{p_\mathrm{out}}$\textbf{[bar]} &
      $\boldsymbol{s}$\textbf{[mm]}\\
      \hline
      {\textit{OP}\,120\,bar} & 0.034  & 120 &  4  & 8.75    \\ 
      {\textit{OP}\,201\,bar} & 0.020  & 201 &  4 & 16.75    \\ 
      \hline
      \end{tabular}
    \label{tab:ops}
  \end{table}

\section{Numerical Setup}\label{s:NumSetup}
  
  We carry out simulations of two representative operating points, which are listed in Table~\ref{tab:ops} and explained in more detail in the following section. Due to the different cavitation intensities, different stand-off distances are chosen for the different operating points. 

  \Cref{fig:setup}~(b) shows the numerical setup. The configuration consists of the inlet chamber, the nozzle and the outflow region confined by the target in the bottom. For the simulations we have simplified the outflow region to a cylindrical domain with the size of the target. The nozzle throat diameter is $d =0.45\,\si{mm}$, the nozzle cavitator expands to about $7 \times d$ and the target to more than $> 100 \times d$. At the inlet the pressure $p_\mathrm{in}$ is prescribed and the circumferential, lateral side of the outflow region is defined as pressure outlet with $p_\mathrm{out}$. All walls are treated as isothermal no-slip walls. The upper boundary of the outflow area is modeled with slip walls, see \cref{fig:setup}~(b).  

  \begin{table}[!tb]
    \centering
    \caption{Data to the different grid levels\new{. The last number of the given resolution on the target refers to the wall-normal direction.} }
      \begin{tabular}{|rrrr|}
      \hline
       \textbf{Grid level [-]} & 
       \textbf{Resolution on target}\textbf{\,[\si{\mu m}]} & 
       \textbf{M cells [-]} & 
        $\boldsymbol{\Delta  t [\mathrm{s}]}$  \\
      \hline
      0  & 90 x 90 x 30      & 0.62   & 4.6e-9  \\
      1  & 45 x 45 x 15      & 4.96   & 2.3e-9  \\
      2  & 22.5 x 22.5 x 7.5 & 39.68  & 1.15e-9 \\
      \hline
      \end{tabular}
    \label{tab:grid}
  \end{table}

  \begin{figure}[!tb]
    \centering
    \subfigure[]{\includegraphics[height=5cm]{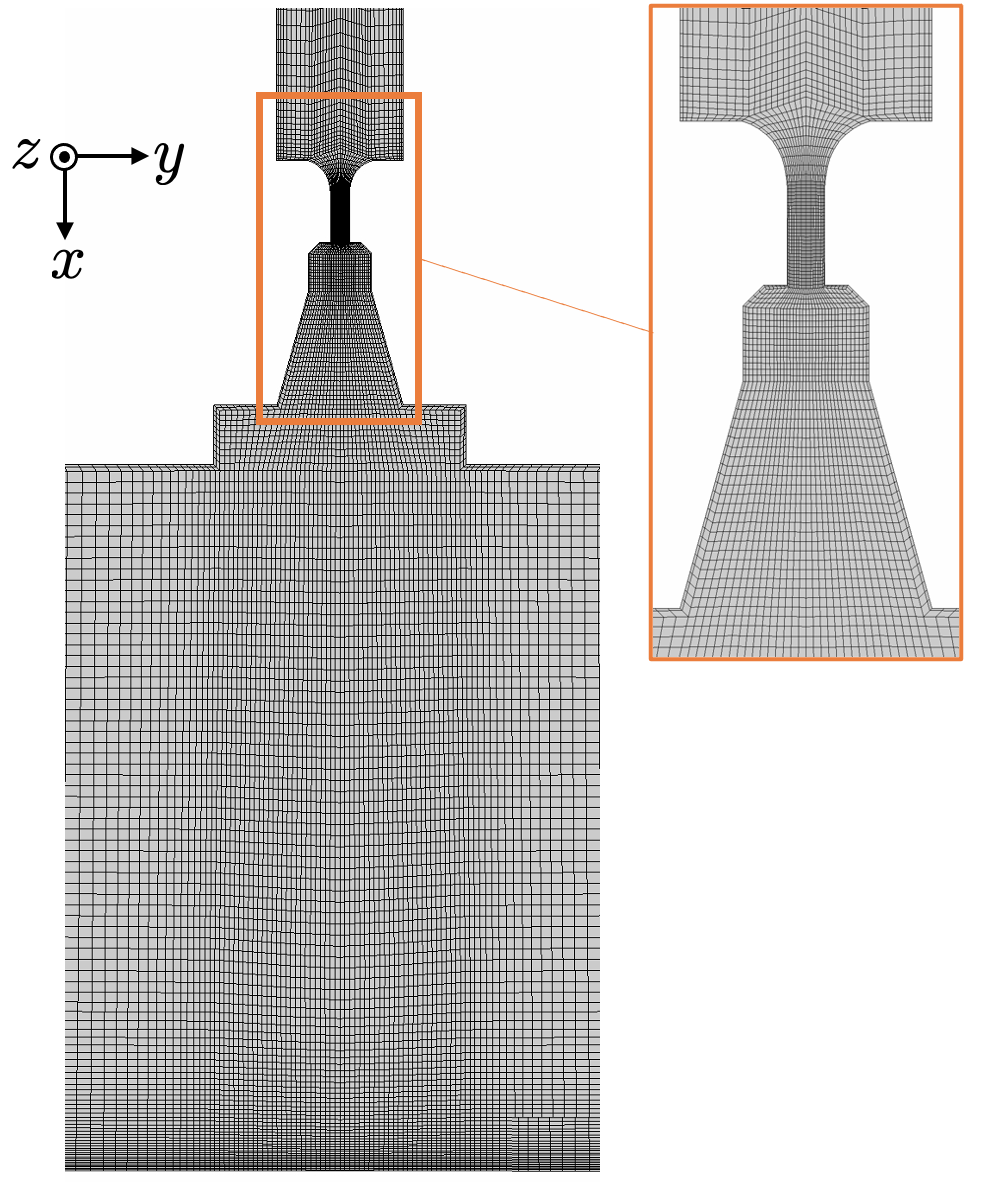}}
    \subfigure[]{\includegraphics[height=5cm]{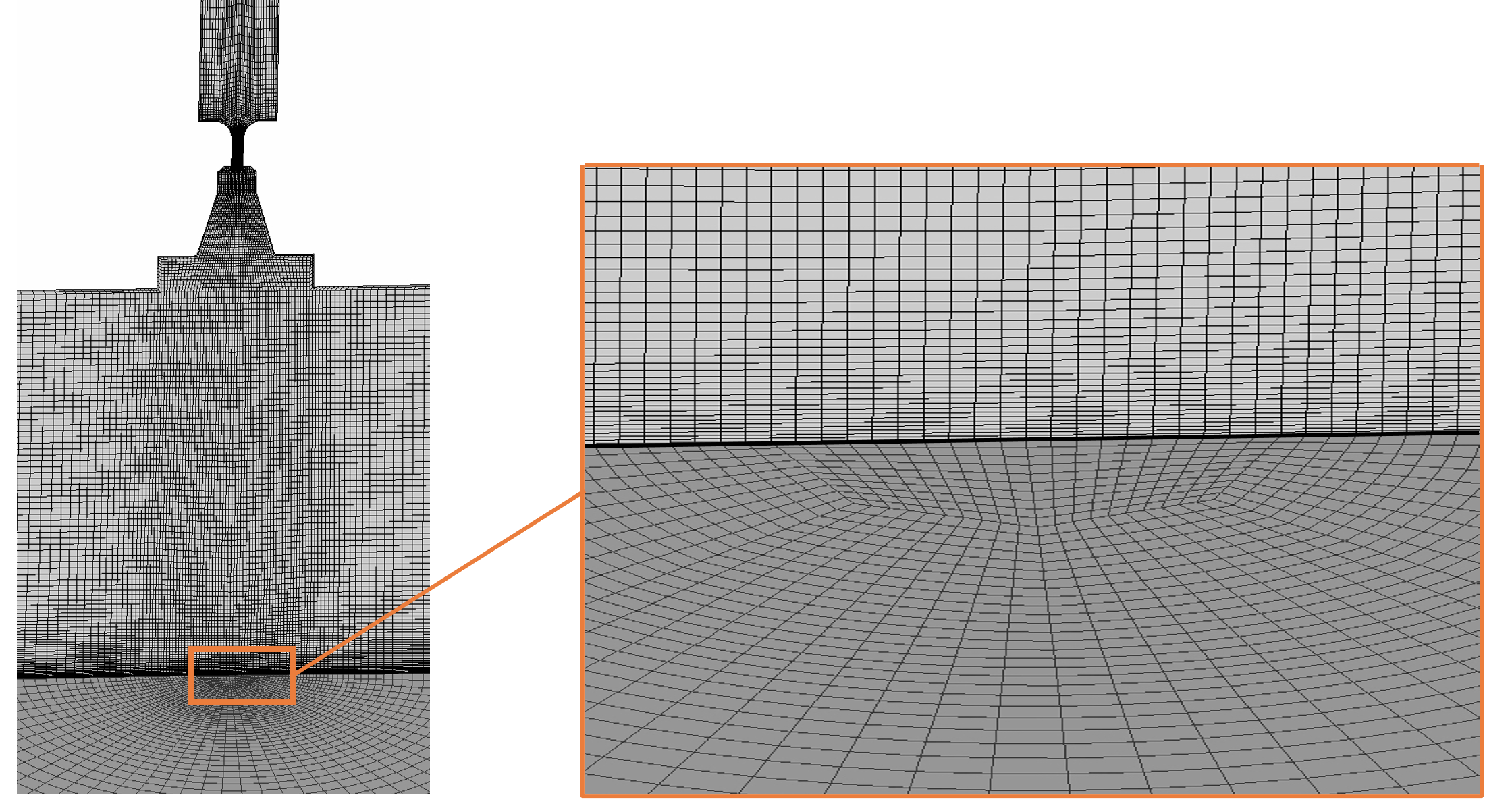}}
    \caption{Grid on the midplane in $x$-$y$ view and zoom of the nozzle region in (a) and grid on the midplane and on the target in 3-D view and zoom of the target region in (b). The coarsest grid resolution is shown to facilitate visibility.}
    \label{fig:grid}
  \end{figure} 

  We discretize the domain using a block-based, structured grid. For the grid sensitivity study, three different grid levels are used \textit{Lvl}~0, \textit{Lvl}~1, \textit{Lvl}~2, see \cref{tab:grid}. To reduce computational costs, the developed flow field on a coarser grid level is used as initial flow field for the next refinement level resulting in a faster convergence \new{to a statistically stationary state}. Refinements between the different grid levels were performed by dividing each cell in every direction. The coarsest grids contain less than a million cells while the finest consist of about 40 million cells. In the finest grids, the smallest cell size in wall-normal direction is 7.5 $\si{\mu m}$. \Cref{fig:grid} depicts slices through the grid. To facilitate visibility, the coarsest grid is shown, meaning that only every fourth grid line of the fine grid is shown. Note that for the two operating points investigated different grids were generated due to the different stand-off distances to the target $s$. 

  We set the Courant-Friedrichs-Lewy number (CFL) to 1.4, resulting in a time step of approximately $\Delta t \approx 0.1\,\si{ns}$ on the finest grid. Statistical averaging on the finest grid was performed over $4\,\si{ms}$, sampling every time step. The analyzing time of $4\,\si{ms}$ corresponds to roughly twenty shedding cycles, depending on the operating point. For the analyses performed, the integral vapor content in the domain and in selected areas such as the diverging nozzle part, the outflow region and the target region was monitored. 

  For our simulations we used the in-house flow solver CATUM (Cavitation at TU Munich), which has been entirely developed by the gas dynamics research group at the chair of Aerodynamics and Fluid mechanics. The simulations were conducted at the SuperMUC and SuperMUC-NG at Leibniz Supercomputing Centre LRZ. On the highest grid resolution simulation ran on about 2000 cores for more than 30 days in total per operating point. 

\section{Results}\label{s:Results}

 In \cref{ss:Overview}, we first provide an overview to flow and cavitation patterns and then we numerically assess the cavitation erosion potential in \cref{ss:Erosion_prediction}. \new{The presented results are obtained on the finest grid resolution (\textit{Lvl 2}).}
  
  \subsection{Flow and cavitation patterns}\label{ss:Overview}

\begin{figure}[!htb]
      \centering
      \subfigure{\includegraphics[width=\textwidth]{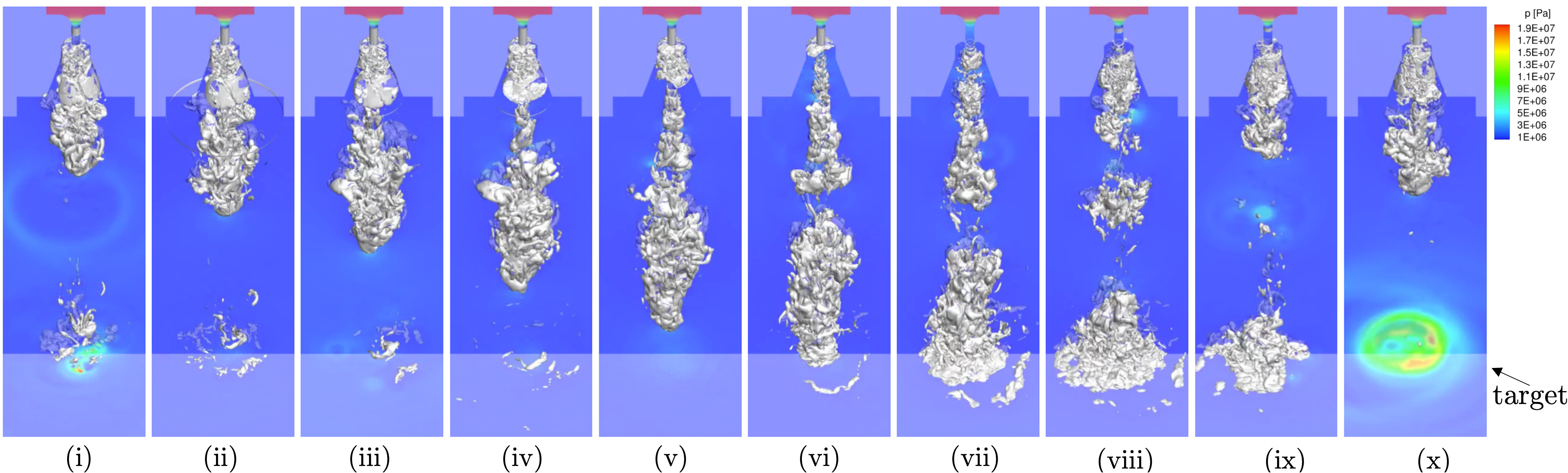}}
      \caption{Time series for \textit{OP} 201 bar visualizing the pressure field and iso-surface vapor (10\%)\new{, the time between the time steps depicted is $24\,\si{\mu s}$. The orthogonal plane in the bottom shows the pressure recorded on the target.}}
      \label{fig:timeseries201}
    \end{figure}

    \Cref{fig:timeseries201} visualizes the flow and cavitation patterns of the cavitating jet for \textit{OP} 201 bar illustrating the dynamic behavior and the main processes. A liquid jet with a velocity of about $200\,\si{m/s}$ emanates in a quiescent liquid leading to shear layer cavitation and subsequently to cloud cavitation with periodic shedding. The shed clouds are convected downstream and finally collapse close to the target (\cref{fig:timeseries201},(x)).

\begin{figure}[!htb]
        \subfigure[]{{\includegraphics[width=0.92\linewidth]{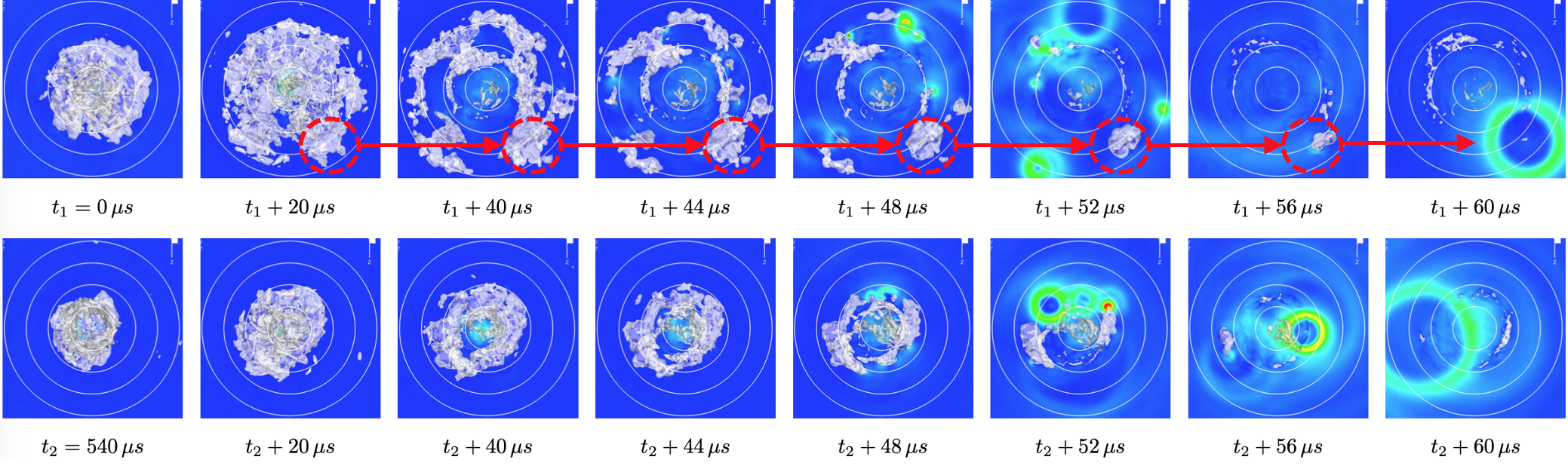}}} 
        \subfigure[]{{\includegraphics[width=0.92\linewidth]{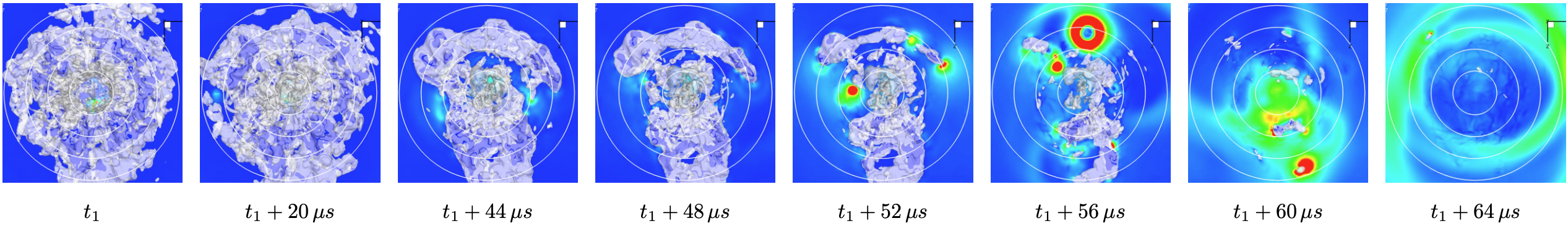}}}        
        \subfigure{{\includegraphics[width=0.06\linewidth]{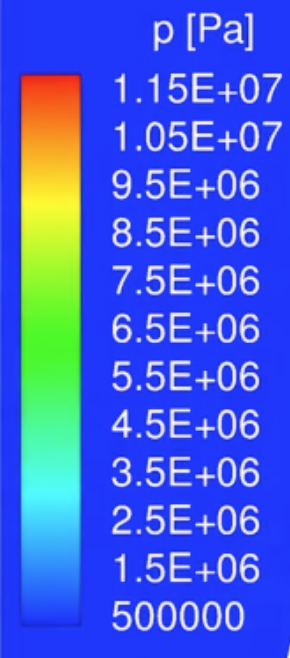}}}\\
      \caption{Time series to visualize erosion process for {\textit{OP}\,120\,bar} (a) and {\textit{OP}\,201\,bar} (b). \new{Top view on the target. Iso-surface vapor (10\%) and pressure on the target are visualized.} The pressure scale was chosen to promote visibility. }
      \label{fig:ts_erosion}
    \end{figure}

    \Cref{fig:ts_erosion} shows time-series of the cavitation and collapse dynamics and the pressure on the target in a top view for {\textit{OP}\,120\,bar} in (a) and {\textit{OP}\,201\,bar} in (b). At the target the vapor cloud expands radially outwards and then fragments. The fragmented vapor structures often have a half-moon shape or are parts of torus segments. These fragments collapse under emission of intense pressure waves. In the time-series, it can be seen that strong collapses mainly take place in the ring-shaped area where the fragmented clouds are convected to. This matches experimental observations on collapse dynamics of a cavitating jet and the ring-shaped erosion pattern, see e.g. \citep{Fujisawa:2017jd, Fujisawa:2019ht}. At {\textit{OP}\,201\,bar} the vapor content in the target is higher and the vapor structures extend further outwards in the radial direction. 

    Note that the collapse dynamics occur on a very small time scale since the time span between maximum vapor content in the target region and collapse of the structures is only about $40 - 60 \,\si{\mu s}$. Based on analyses of the recorded vapor signals (see e.g. \cref{fig:corr_vap_co_ev,fig:grid_vapor}), we found an average shedding period of $168\,\si{ns}$ and $200\,\si{ns}$ for {\textit{OP}\,120\,bar} and {\textit{OP}\,201\,bar}, respectively. 

    In the subsequent section, we numerically assess the erosion potential induced by the collapsing vapor structures. 

\FloatBarrier 

       \begin{figure}[!htb]
      \centering
      {\includegraphics[width=0.8\linewidth, trim={0 0 0 0},clip]{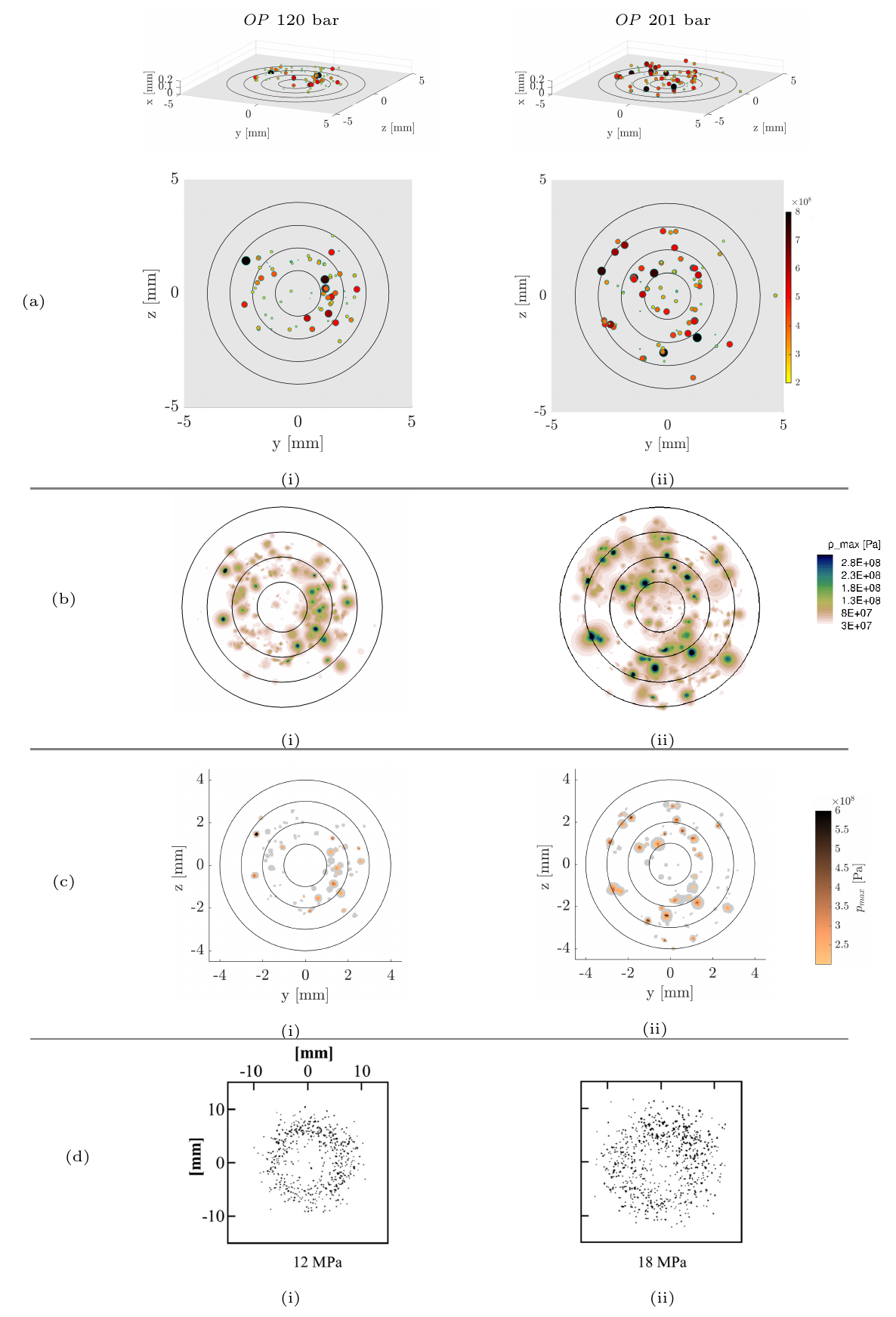}}
      \caption{Comparison of the erosion potential of {\textit{OP}\,120\,bar} (left column) (i) and {\textit{OP}\,201\,bar} (right column) (ii). 
      (a) (top panel): detected collapse events in the near target region ($\Delta x= 200 \, \mu m$), color and size correspond to the recorded collapse pressure $p_c$, only collapses with $p_c>\num{2e8}\,\si{Pa}$ are shown. 
      (b) (second panel): recorded maximum wall pressure in the simulations. 
      (c) (third panel): cell-based pressure data for generating NPEs (numerical pit equivalents), colored data $p>\num{2e8}\,\si{Pa}$, data in light grey $\num{2e8}\,\si{Pa}>p>\num{1e8} \,\si{Pa}$. 
      All simulation data for an analyzing time of $\Delta t=4\,\si{ms}$, orientation rings at $r=1,2,3,4\,\si{mm}$. 
      (d) (bottom panel): Experimental results by \citet{Fujisawa:2019ht} for a comparable configuration but different inlet pressures, different geometry of the setup, and a longer exposure time (reprinted with permission granted by Elsevier). }
      \label{fig:erosion_pmax}
    \end{figure}%
\FloatBarrier

  \subsection{Numerical prediction of cavitation erosion}
  \label{ss:Erosion_prediction}

    In this section, we present and compare different approaches for numerical cavitation erosion prediction and introduce numerical pit equivalents (NPEs). For all evaluations, we used a data set covering $4\,\si{ms}$ which corresponds to 24 and 20 shedding cycles of {\textit{OP}\,120\,bar} and {\textit{OP}\,201\,bar}, respectively. 

    \subsubsection{Collapse detection}
    Cavitation-induced material damage is associated with violent collapses of vapor clouds emitting intense pressure waves, see \cref{fig:timeseries201,fig:ts_erosion}. To capture such collapse events in our numerical simulations, we utilize the collapse detection algorithm developed by \citet{Mihatsch:2015db}. Therewith collapse events are identified based on vanishing vapor content in a cell and all next-neighbor cells and a negative divergence of the velocity. The maximum pressure is recorded and stored as collapse pressure $p_{c}$. \Cref{fig:erosion_pmax}~(a) visualizes the detected collapse events in the near target region ($\Delta x= 200 \, \mu m$) where color and size correspond to the recorded collapse pressure. More collapses with high collapse pressures are detected at \textit{OP} 201 bar. Further, at {\textit{OP}\,201\,bar} the distribution of the collapses expands further outwards in the radial direction, which correlates with the distribution of the vapor structures in \cref{fig:ts_erosion}. 

        \begin{figure}[!tb]
      \centering
        \subfigure[]{\includegraphics[width=0.35\linewidth]{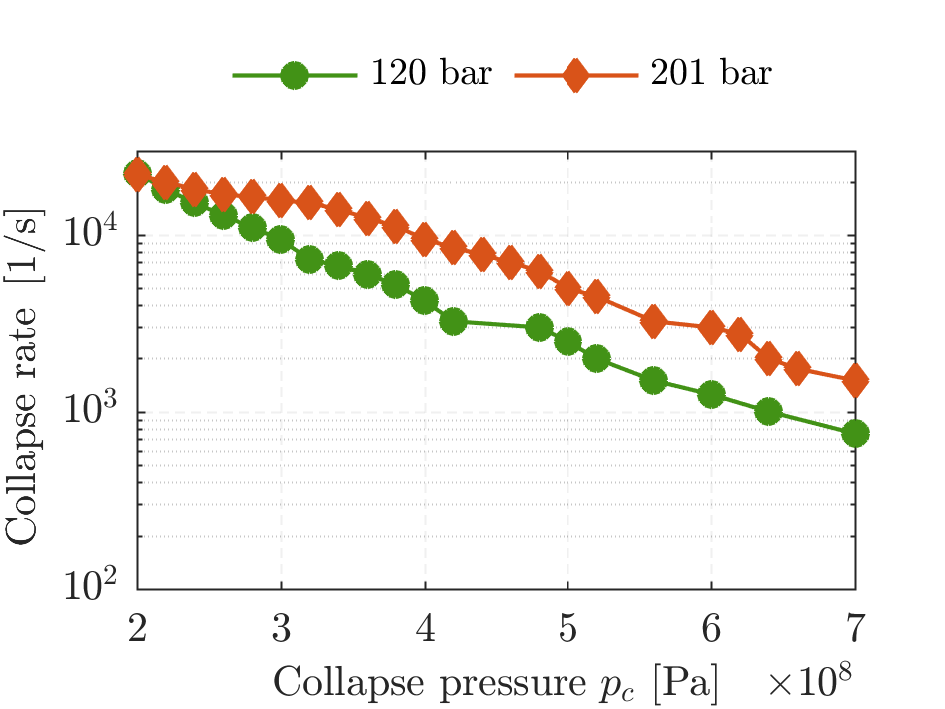}}
        \subfigure[]{{\includegraphics[width=0.35\linewidth]{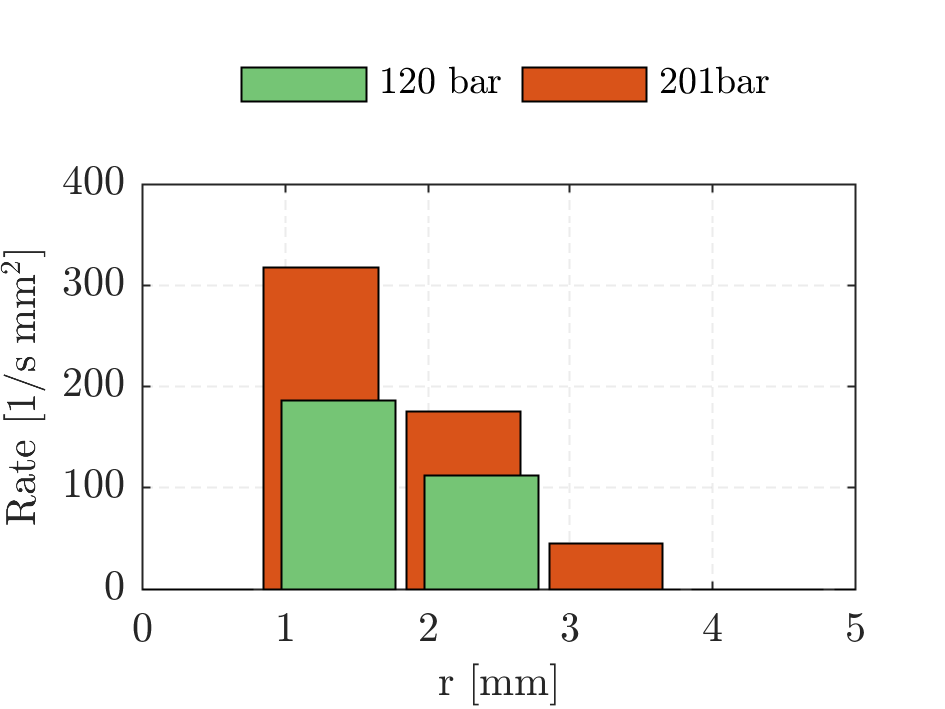}}}
        \subfigure[]{{\includegraphics[width=0.35\linewidth]{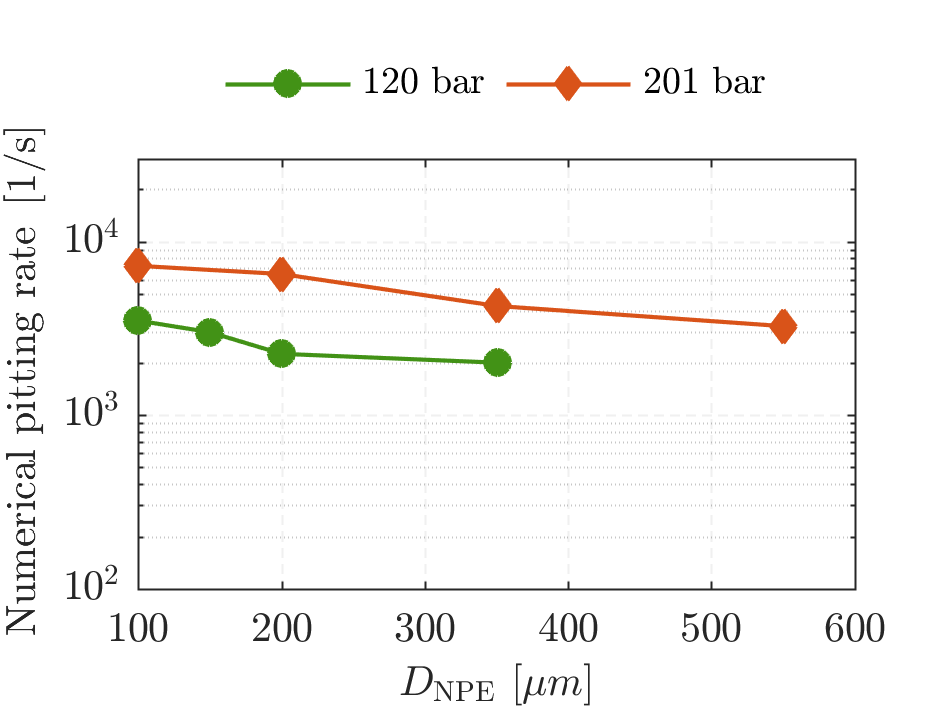}}}
        \subfigure[]{{\includegraphics[width=0.35\linewidth]{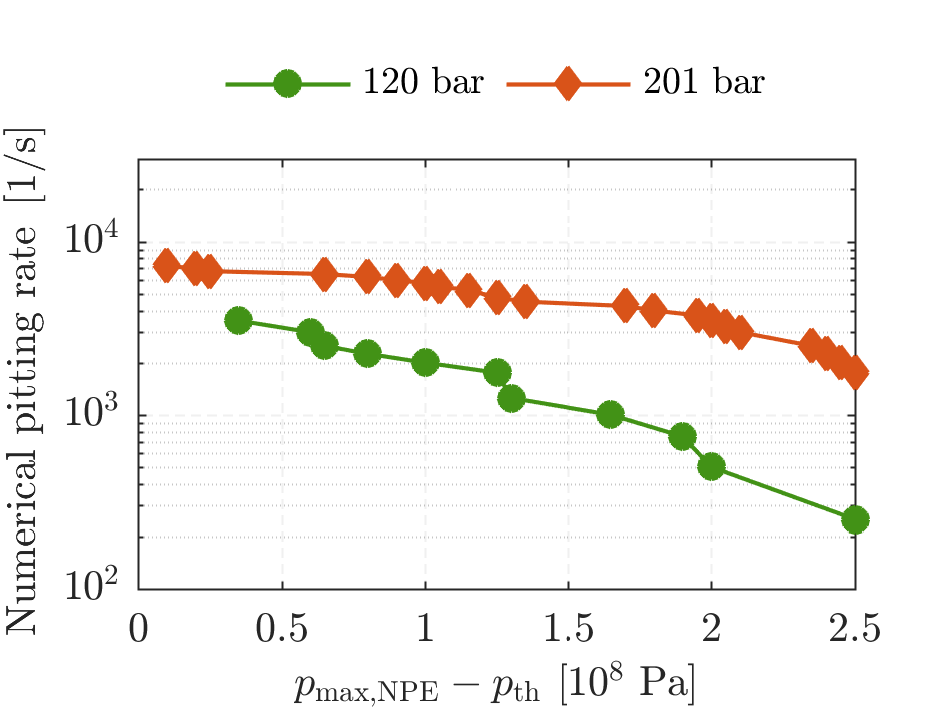}}}
      \caption{Cumulative collapse spectra in (a), numerically obtained pitting rates in (b) (Note that the reference area increases in the radial direction ($A_{i \to i+1}=(r_{i+1}^2-r_{i}^2)\pi$)), cumulative numerical pit spectra over $D_\mathrm{NPE}$ in (c) \new{and over the maximum pressure above threshold ($p_\mathrm{max,NPE}-p_\mathrm{th}$) (which is associated with the pit depth $\delta_\mathrm{NPE}$) in (d).} All data generated using $p_\mathrm{th}=\num{2e8}\,\si{Pa}$. }
      \label{fig:spectra}
    \end{figure} 

    Using the collapse data also cumulative collapse spectra can be generated, see \citet{Mihatsch:2015db}. \Cref{fig:spectra}~(a) shows the cumulative collapse spectra of the detected collapse events in the near target region ($\Delta x= 200 \, \mu m$) with a collapse pressure $p_c>\num{2e8}\,\si{Pa}$. The higher cumulative collapse rates at higher pressures at {\textit{OP}\,201\,bar} indicate a higher erosion aggressiveness of this OP. While the collapse detection algorithm allows for the assessment of the collapse dynamics~\citep{Egerer:2014wu,Orley:2015kt} and for quantitative comparison of different operating points, geometries~\citep{Beban:2017vo} and phases during an injection cycle~\citep{Orley:2016db}, a direct estimation of potential cavitation-induced material damage is not possible. However, these can be e.g. approximated based on $p_c$ and the distance of the collapse to the wall using \new{the correlation between the maximum pressure and distance to the focus point of the collapse, }see \citet{Mihatsch:2017diss}. 

    \subsubsection{Maximum wall pressure}
    For the actual identification of surface areas prone to erosion damage, a straightforward approach is the evaluation of surface loads identified by pressure maxima on the wall. In several previous studies of e.g. channel flows~\citep{Egerer:2014wu}, throttle valves~\citep{Beban:2017vo}, and for single bubble collapses~(e.g.~\citet{Lauer:2012jh,Pishchalnikov:2018pp,Trummler:2021if, Trummler:2020JFM}) this quantity has been successfully employed for numerical estimations of the erosion potential. \Cref{fig:erosion_pmax}~(c) shows the maximum pressure recorded in the simulations. As expected, the distribution of the pressure maxima shows a strong spatial correlation with the detected collapse events. At {\textit{OP}\,120\,bar} most of the pressure impacts on the wall are between $1-3\,\si{mm}$ while at {\textit{OP}\,201\,bar} these are mainly scattered between $1-4\,\si{mm}$ and more intense, i.e. they have a higher pressure maxima at the center and they are larger. The increase of the radial extent with increasing inlet pressure is in good agreement with experimental findings for comparable configurations. In \cref{fig:erosion_pmax}~(d), experimental data from a comparable configuration \citep{ Fujisawa:2019ht} are shown. Considering the fact that the sampling time and the geometrical dimensions differ, the results are in good qualitative agreement. 

    \subsubsection{Numerical pit equivalents (NPEs)}
    While the maximum wall pressure is easy to visualize and gives a good impression of the regions prone to erosion, further data processing is required to obtain more quantitative information. For a better prediction of the potential wall damage and better comparability with experimental data, we suggest generating numerical pit equivalents (NPEs). 
    
    \new{The evaluation method works as follows: W}e monitor and output the hydrodynamic pressure exceeding a certain threshold value in the near-wall cells\new{. When the threshold value is exceeded, the start time of the pressure event is recorded, and the evaluation of the peak pressure is started. When the pressure falls below the threshold value, the duration and the peak pressure are evaluated and written out, representing the cell-based pressure data. This approach can be extended to additional quantities that can serve as erosion indicators, such as the pressure impulse $\int (p -p_\mathrm{th})\, \mathrm{d} t$ (see e.g. \citet{schmidt2014assessment,Trummler:2020JFM}). Furthermore, it is worth noting that only the duration and integral quantities depend on the chosen threshold, but pressure peak values can easily be filtered a-posteriori using a higher threshold. T}o generate NPEs \new{from the cell-based pressure data,} we employ a physically motivated clustering algorithm to recognize spatially and temporally contiguous areas featuring high pressure~\citep{Trummler:2021CAV}. Therewith we obtain NPEs characterized by their peak pressure $p_\mathrm{max,NPE}$, their spatial distribution, e.g. their diameter $D_\mathrm{NPE}$, and the duration of their formation $\Delta t_\mathrm{NPE}$. 

    In the considered application, we employ the two threshold values of $p_\mathrm{th} = \num{1e8} \,\si{Pa}$ and $p_\mathrm{th} = \num{2e8} \,\si{Pa}$\new{, where t}he latter roughly corresponds to the ultimate tensile strength of copper with $\num{2.1e8} \,\si{Pa}$ \citep{ross2013metallic}. \new{These threshold values are empirical simulation parameters, and $p_\mathrm{th} = \num{2e8} \,\si{Pa}$ has been chosen to filter out those events that can potentially cause deformation of the copper target. More appropriate threshold values could be defined using material properties determined by nanoindentation tests that better represent pit formation~\citep{Roy:2015jx,Roy:2015koa,Roy:2015cl,Fivel:2015dd,carnelli2012evaluation}. Further, the small time scales of collapse events result in high strain rates affecting the material response~\citep{Roy:2015koa,Fivel:2015dd,joshi2020bubble}, thus, data from compressive tests at high strain rate should ideally be considered.}

\begin{figure}
      \centering
         \subfigure{\includegraphics[trim={0 0 0 0}, clip, height=12cm]{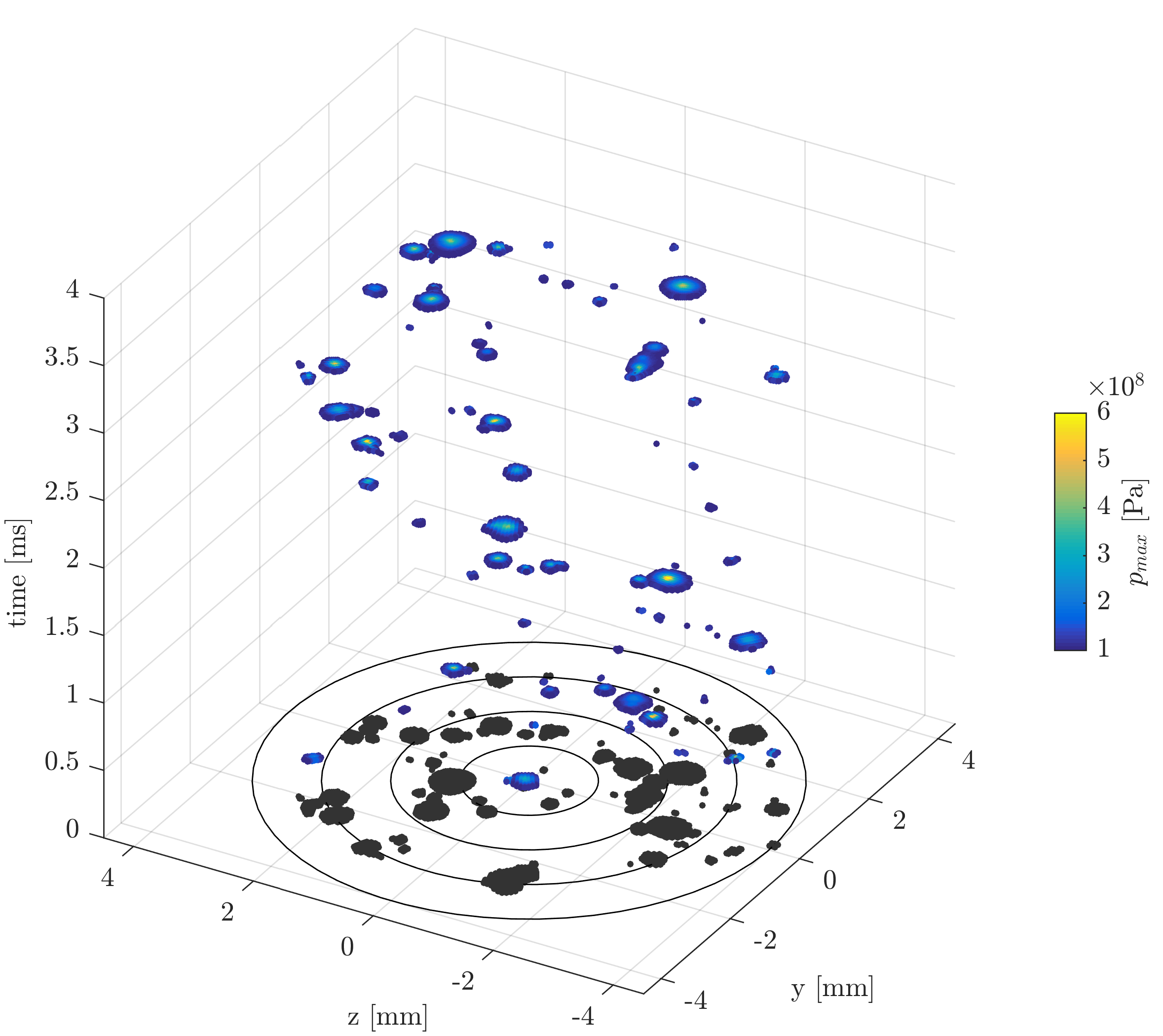}}
         \caption{\new{Cell-based pressure data used to generate NPEs for {\textit{OP}\,201\,bar}. The dark gray data in the bottom show all pressure data recorded over $\Delta t = 4\,\si{ms}$ and match those in \cref{fig:erosion_pmax}~(c)}. } 
        \label{fig:NPEs_time}
    \end{figure}

    \new{The cell-based pressure data is visualized in \cref{fig:erosion_pmax}~(c). This presentation \newnew{yields} the same results as the maximum wall pressure (\cref{fig:erosion_pmax}~(b)), however, it should be noted that this data also include time information as depicted in \cref{fig:NPEs_time} for {\textit{OP}\,201\,bar}. The additional time information allows for the distinction of pressure impacts that are spatially, or temporally, close to each other, such as overlapping pits. Moreover, also pitting rates can be determined, as presented in the following.} 

\FloatBarrier

    \begin{figure}
      \centering
         \new{\subfigure[]{\includegraphics[width=0.8\linewidth]{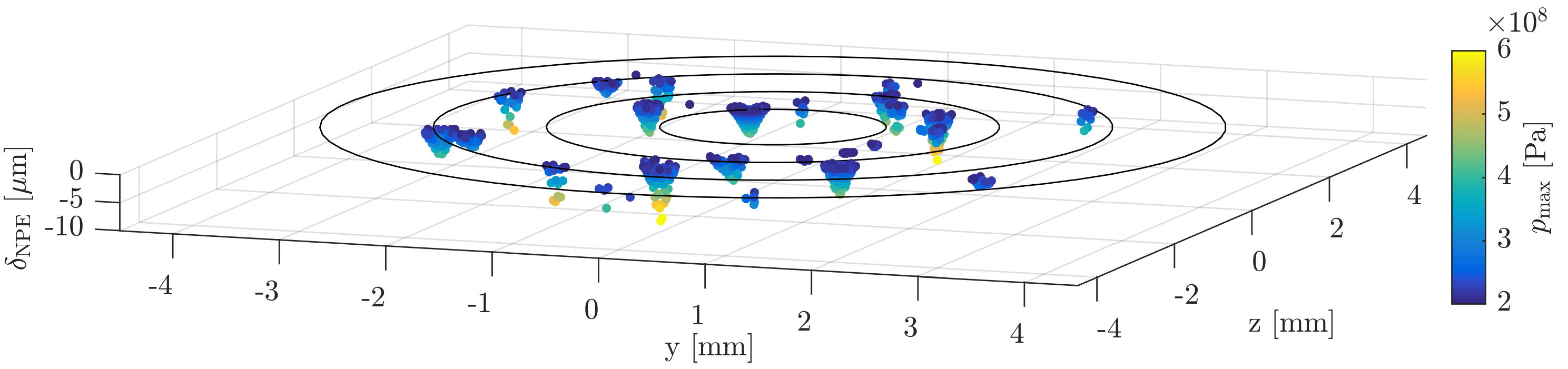}}}\\
         \subfigure[]{\includegraphics[height=5cm]{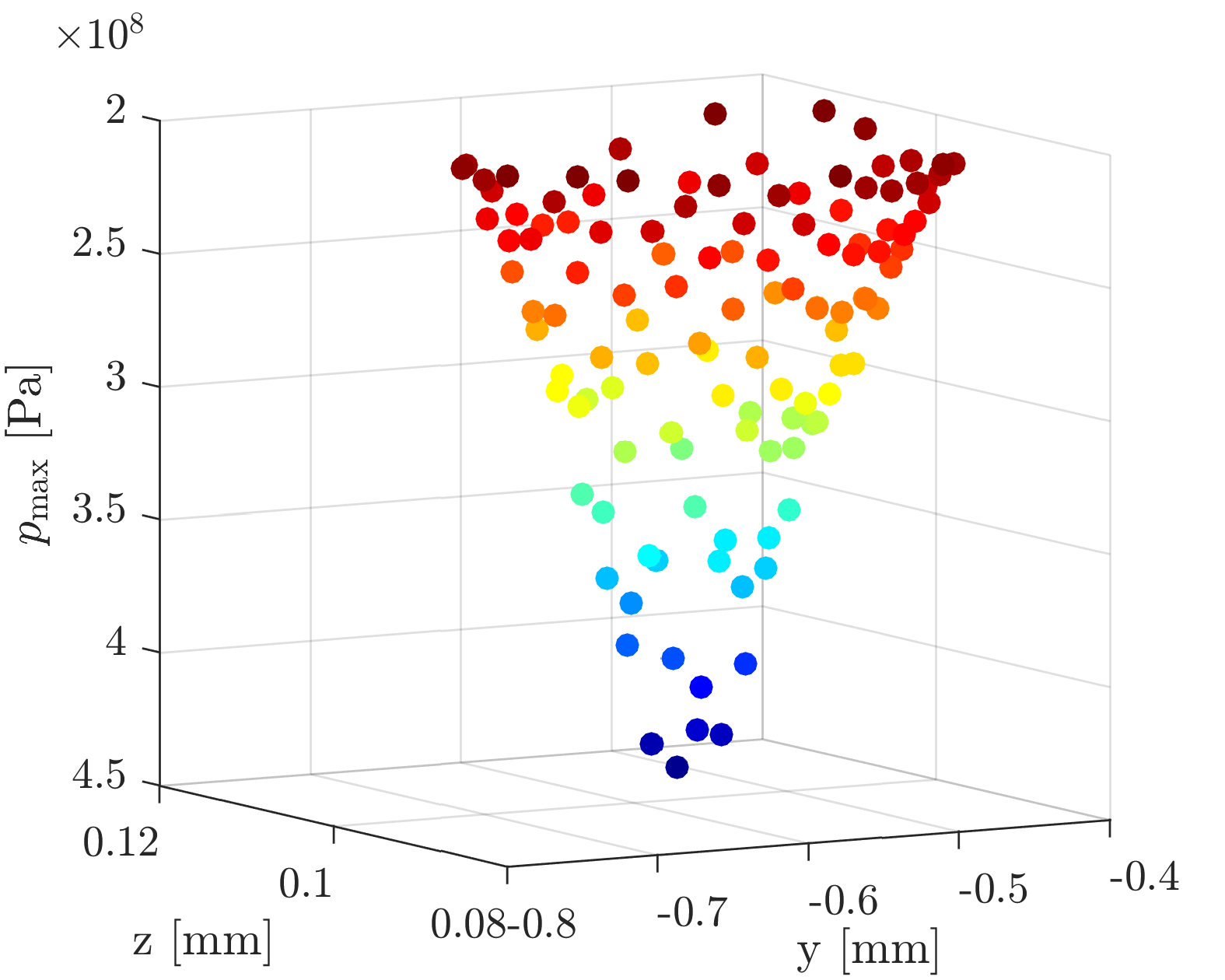}}
         \subfigure[]{\includegraphics[height=5cm]{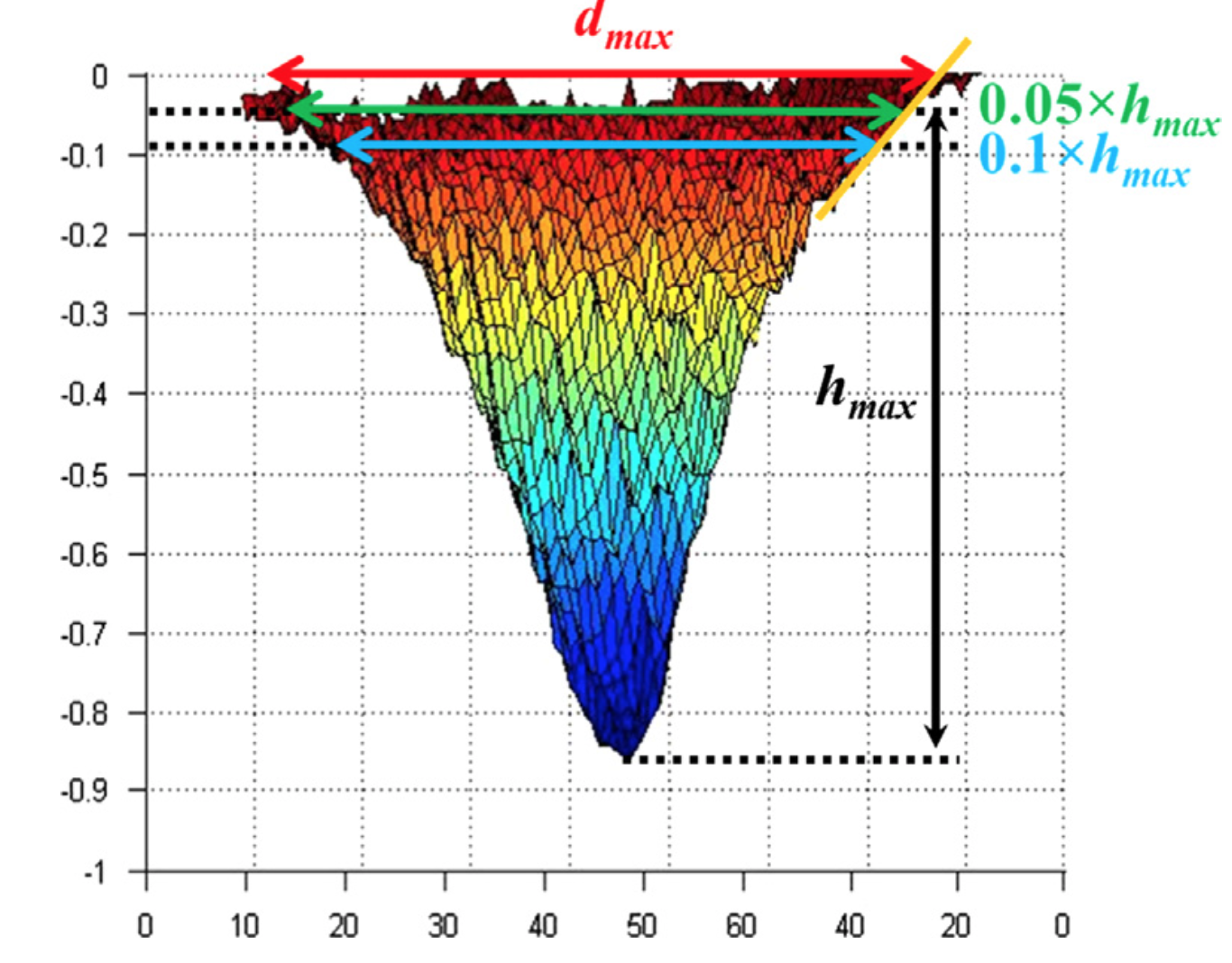}}
         \new{\subfigure[]{\includegraphics[height=4cm]{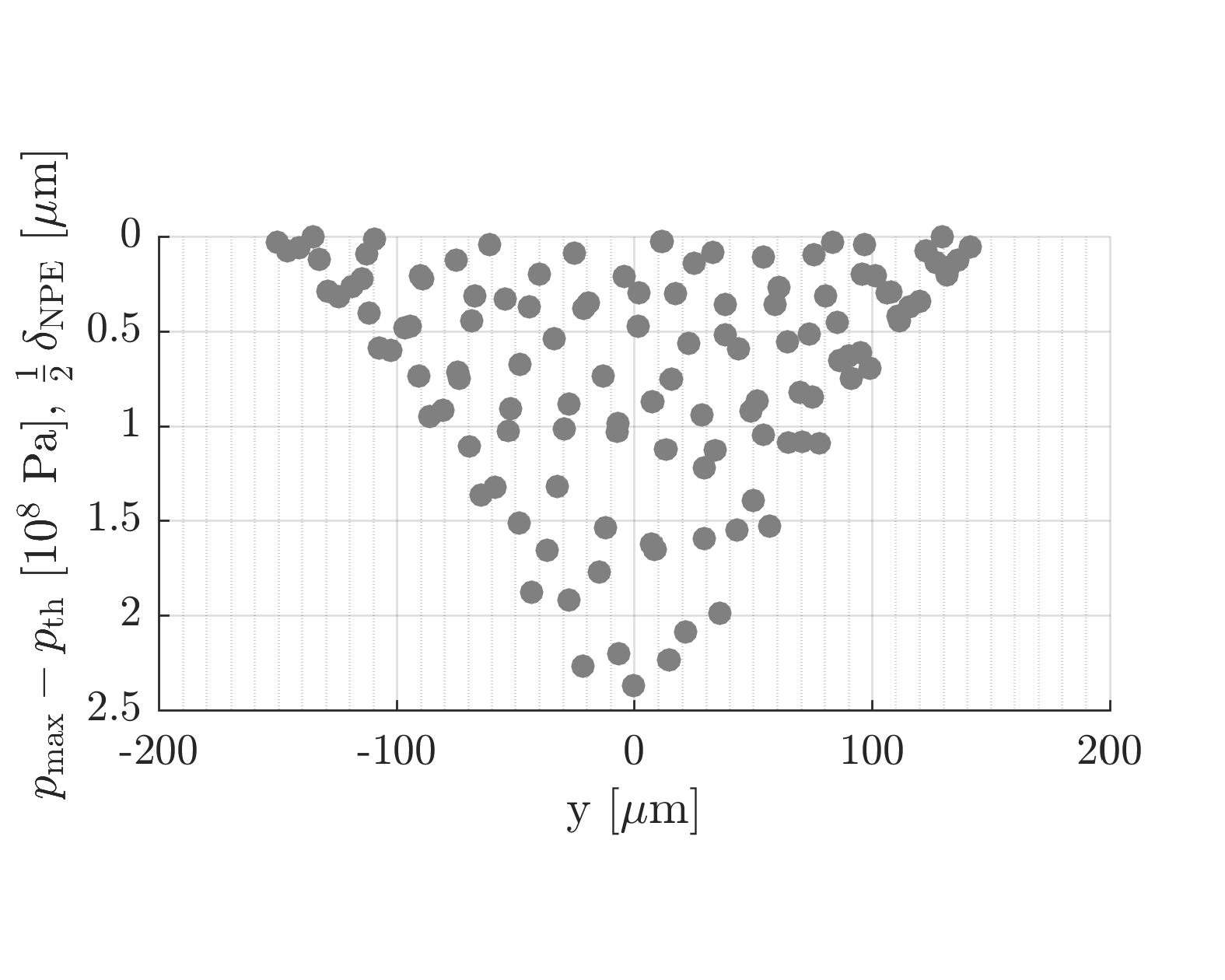}}}
         \new{\subfigure[]{\includegraphics[height=5cm]{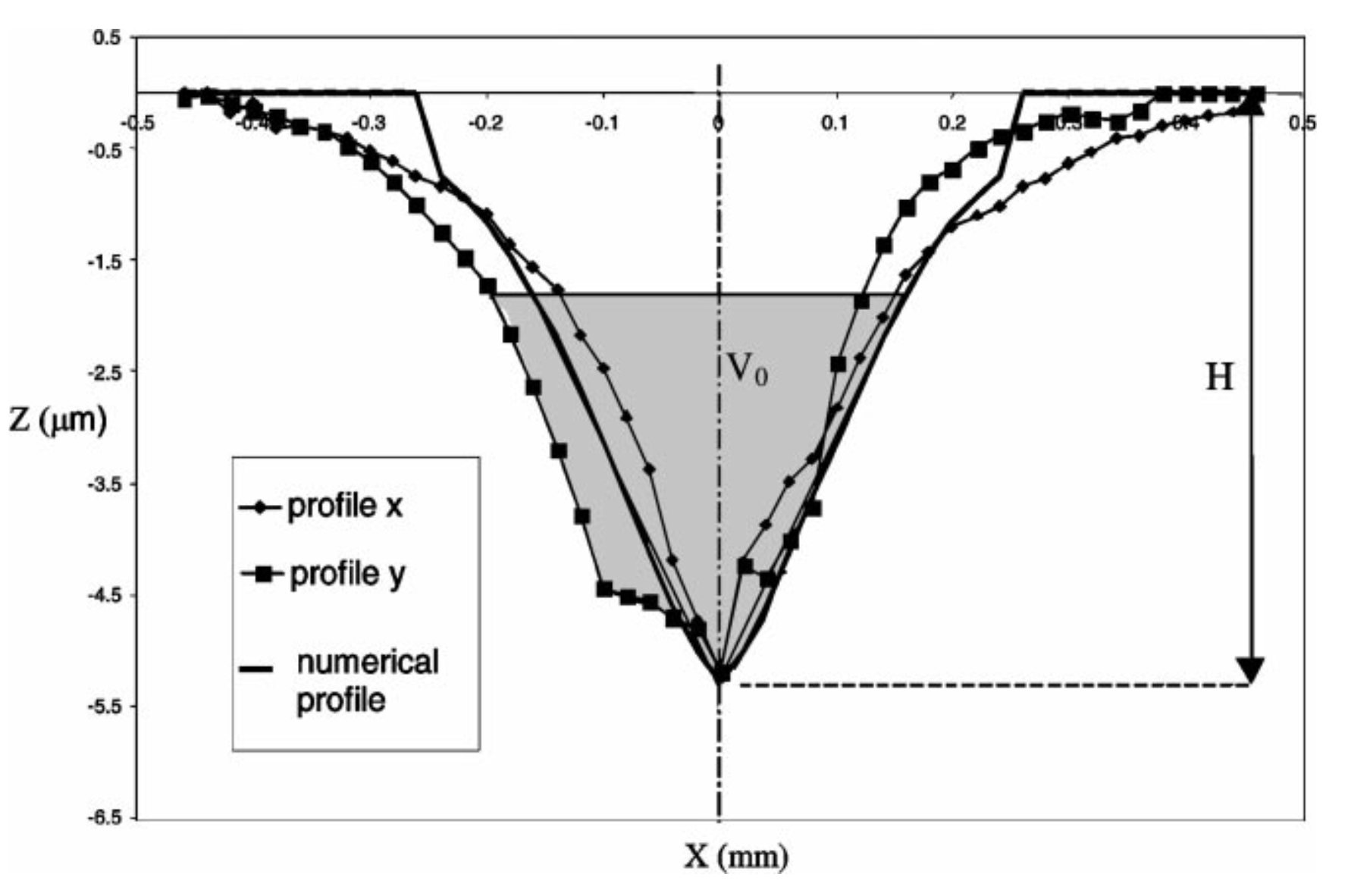}}}
         \caption{\new{Visualization of NPEs and comparison with experimental data. (a)~All data for {\textit{OP}\,201\,bar} (pit depth $\delta$ is estimated using \cref{eq:delta}, (b)}~distribution of maximum wall pressure $p_\mathrm{max}$ of one selected pit in our simulation \new{(each point represents the value recorded in a computational cell), (c)~}experimental results of \citet{Roy:2015jx} \new{measured on an aluminum sample in a cavitation loop, shown \newnew{in} a dimensionless length scale, for details see reference} (reprinted with permission granted by Elsevier)\new{, (d) selected pit in our simulation in a frontal view, (e) example of a pit profile observed by \citet{patella2000cavitation} on a copper sample surface (the indentation profile is modeled using an axisymmetric approach, $V_0$ denotes a reference volume evaluated from $H$ to $1/3 H$, the samples were damaged in wake cavitation, for details to the experiment see reference) (reprinted with permission granted by Elsevier).}
         } 
        \label{fig:pit}
    \end{figure}

    Generating NPEs allows for versatile analyses as well as better comparability with experimental ones. First of all, \new{equivalents for experimental pit diameters and pit depths can be estimated, and secondly, pit numbers and pitting rates can be evaluated.} For the sake of simplicity, we take for the pit diameter $D_\mathrm{NPE}$ the mean diameter of the area exceeding the threshold $p_\mathrm{th}$. \new{We here numerically predict} pit diameters of about $D_\mathrm{NPE}=100-600\,\si{\mu m}$ for $p_\mathrm{th}=\num{2e8}\,\si{Pa}$\new{. For actual predictions of pit sizes, more sophisticated modelling of the material behavior is required. T}he pit depth $\delta_\mathrm{NPE}$ \new{can be estimated using a functional correlation between $p_\mathrm{max,NPE}$ and the material deformation and by fitting the constants with suitable experimental data. }For simplicity, we employ a linear correlation in the form of 
    \begin{equation}
      \delta_\mathrm{NPE}= k_\mathrm{mat}\cdot(p_\mathrm{max,NPE}-p_\mathrm{th}),
      \label{eq:delta}
      \end{equation} 
    where we fitted $k_\mathrm{mat}=\num{2e-14}\,\si{m/Pa}$ to \new{roughly} match experimentally \new{determined pit depths for copper~\citep{patella2000cavitation}. \Cref{fig:pit}~(a) shows all NPEs in 3-D visualization. In \cref{fig:pit}~(b-d), an NPE from our simulations is compared with experimental pit data from the literature~\citep{Roy:2015jx,patella2000cavitation}, see caption of \cref{fig:pit} for details. The spatial distribution of the maximum wall pressure in our simulation agrees with the plastic deformation measured in the experiments.}

   \FloatBarrier

   To demonstrate the versatile analysis possibilities using NPEs, we have evaluated the pitting rates and cumulative rates over the estimated pit diameters and \new{over the maximum pressure above threshold ($p_\mathrm{max,NPE}-p_\mathrm{th}$), which is associated with the pit depth $\delta_\mathrm{NPE}$.} The numerically obtained pitting rates at different radial positions are plotted in \cref{fig:spectra}~(b) and indicate the higher erosion potential at {\textit{OP}\,201\,bar} and the larger radial extent of the area prone to erosion at this operating point. Inspired by the experimental analyses by \citet{franc2012material}, we have also evaluated cumulative rates over pit diameters and \new{the maximum pressure above threshold}. The obtained cumulative rates are depicted in \cref{fig:spectra}~(c,d) and again predict a higher erosion potential of \textit{OP}\,201\,bar. For all evaluations shown in \cref{fig:spectra}, we used $p_\mathrm{th}=\num{2e8} \,\si{Pa}$\new{.} 
 
  \FloatBarrier  
   By monitoring the hydrodynamic pressure in the near-wall cells, we also obtain the duration of the formation of the NPEs $\Delta t_\mathrm{NPE}$. For $p_\mathrm{th}=\num{2e8}\, \si{Pa}$, the duration is about 20 to 100 $\si{ns}$, which correlates with the characteristic time scale of experimentally measured pressure pulses of roughly $<100\,\si{ns}$ \citep{Tinguely:2012wo}. One advantage of high-resolution numerical simulations is the ability to assess processes in detail on small time and length scales, such as the temporal and spatial evolution of an NPE. 

   \begin{figure}[!tb]
      \centering
        \subfigure{{\includegraphics[width=0.99\linewidth]{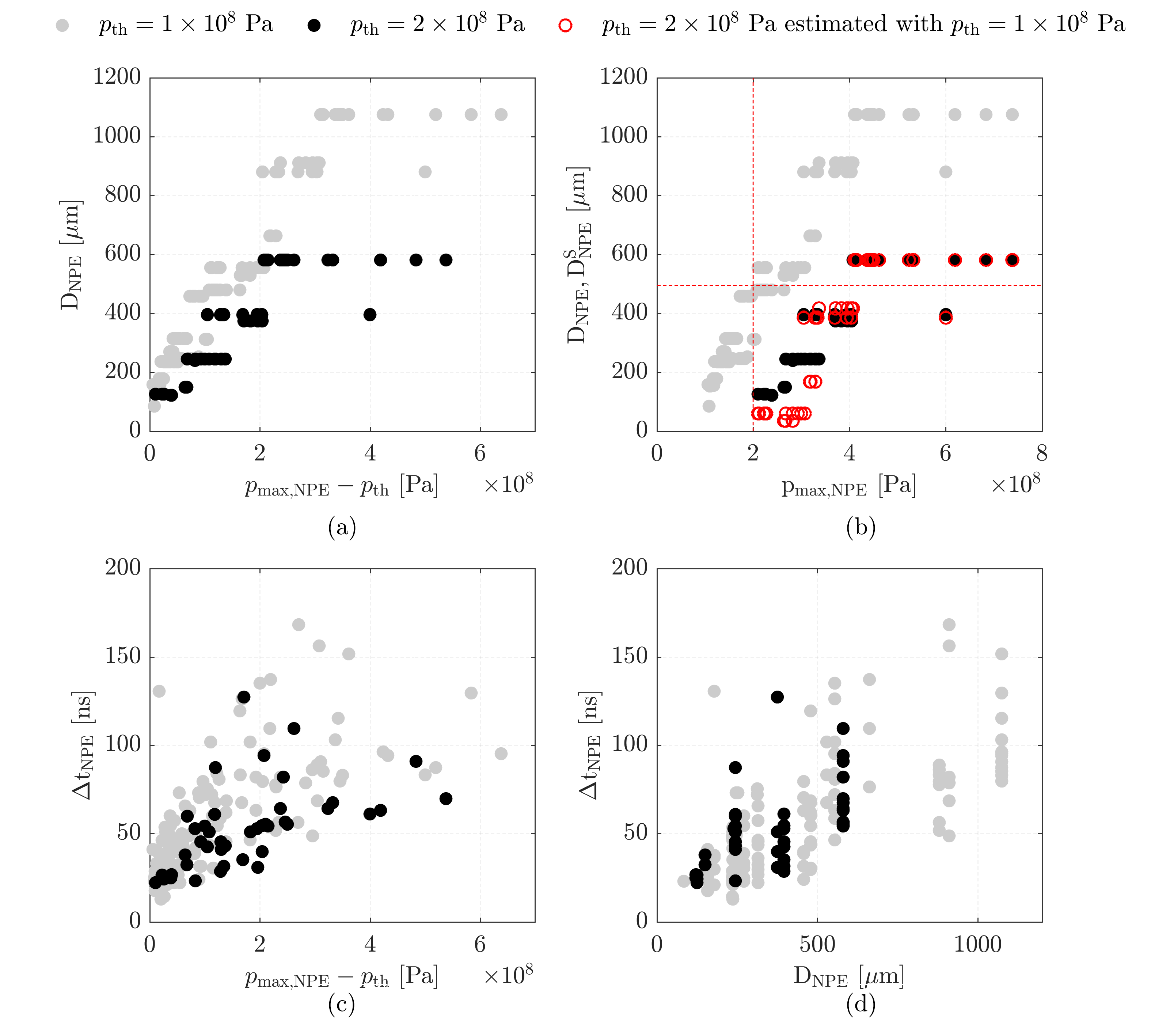}}}
      \caption{Correlations of NPEs for different threshold values $p_\mathrm{th}=\num{1e8}\,\si{Pa}$ and $p_\mathrm{th}=\num{2e8}\,\si{Pa}$, where the latter is erosion-relevant for copper. (a) Correlation of NPE diameter ($D_\mathrm{NPE}$) with maximum pressure above threshold ($p_\mathrm{max,NPE}-p_\mathrm{th}$), (b) NPE diameter ($D_\mathrm{NPE}$) with maximum pressure ($p_\mathrm{max,NPE}$) including the estimation of the diameter $D^S_{NPE}$ for $p_\mathrm{th}=\num{2e8}\,\si{Pa}$ (obtained  with the data of $p_\mathrm{th}=\num{1e8}\,\si{Pa}$ and subtracting the value at $p_\mathrm{th}=\num{2e8}\,\si{Pa}$), (c) duration of pit formation ($\Delta t_\mathrm{NPE}$) with maximum pressure above threshold ($p_\mathrm{max,NPE}-p_\mathrm{th}$), (d) duration ($\Delta t_\mathrm{NPE}$) with diameter ($D_\mathrm{NPE}$).
      }
      \label{fig:correlations}
   \end{figure}
     
     \Cref{fig:correlations} plots the correlation of the parameters of the NPEs $p_\mathrm{max,NPE}-p_\mathrm{th}$, $D_\mathrm{NPE}$, and $\Delta t_\mathrm{NPE}$ including the data for both threshold values. The diameter $D_\mathrm{NPE}$ increases with the maximum pressure exceeding the threshold $p_\mathrm{max,NPE}-p_\mathrm{th}$, which matches experimental observations for the correlation of pit depth and pit diameter~\citep{franc2012material}. We observe a nearly linear increase until an upper limit is reached. \newnew{We assume that this upper limit is related to the potential energy of the largest collapsing vapor structures, which determines the maximum shock wave energy. The latter can be assumed to be approximately proportional to $ d^2 {p_{\mathrm{max},d}}^2$, where $d$ denotes the distance to the focus point. This assumption is supported by the fact that the upper limit is indirectly proportional to the threshold value.} The visible clustering of data points at some diameters is due to the cell size dictating the minimum diameter and the steps between estimated diameters. For many applications, it could be interesting to estimate NPE diameters for different, respectively higher, threshold values. \new{Since there is a nearly linear correlation of $D_\mathrm{NPE}$ and $p_\mathrm{max,NPE}-p_\mathrm{th}$, we suggest the following scaling approach for an estimation of the pit diameter at a threshold value of interest ($\mathrm{tvoi}$):
     \begin{equation}
             D_\mathrm{NPE}^{S,\mathrm{tvoi}} = D_\mathrm{NPE}-D_\mathrm{NPE}(\mathrm{tvoi}),
     \end{equation}
     where the value at which the $\mathrm{tvoi}$ intersects with the current data ($D_\mathrm{NPE}(\mathrm{tvoi})$) is subtracted. \newnew{\Cref{fig:correlations}~(b)} compares the pit diameters obtained for $p_\mathrm{th}=\num{2e8}\,\si{Pa}$ with the ones obtained for $p_\mathrm{th}=\num{1e8}\,\si{Pa}$ and scaled to $\num{2e8}\,\si{Pa}$ ($D_\mathrm{NPE}^{S}$) and shows a good agreement.}

     The duration $\Delta t_\mathrm{NPE}$ increases with the maximum pressure $p_\mathrm{max,NPE}$ (\cref{fig:correlations}~(c)) and the diameter $D_\mathrm{NPE}$ (\cref{fig:correlations}~(d)), \newnew{where a linear correlation may be used to roughly describe the data.}

 \FloatBarrier   

    \begin{figure}[!tb]
        \includegraphics[width=\linewidth]{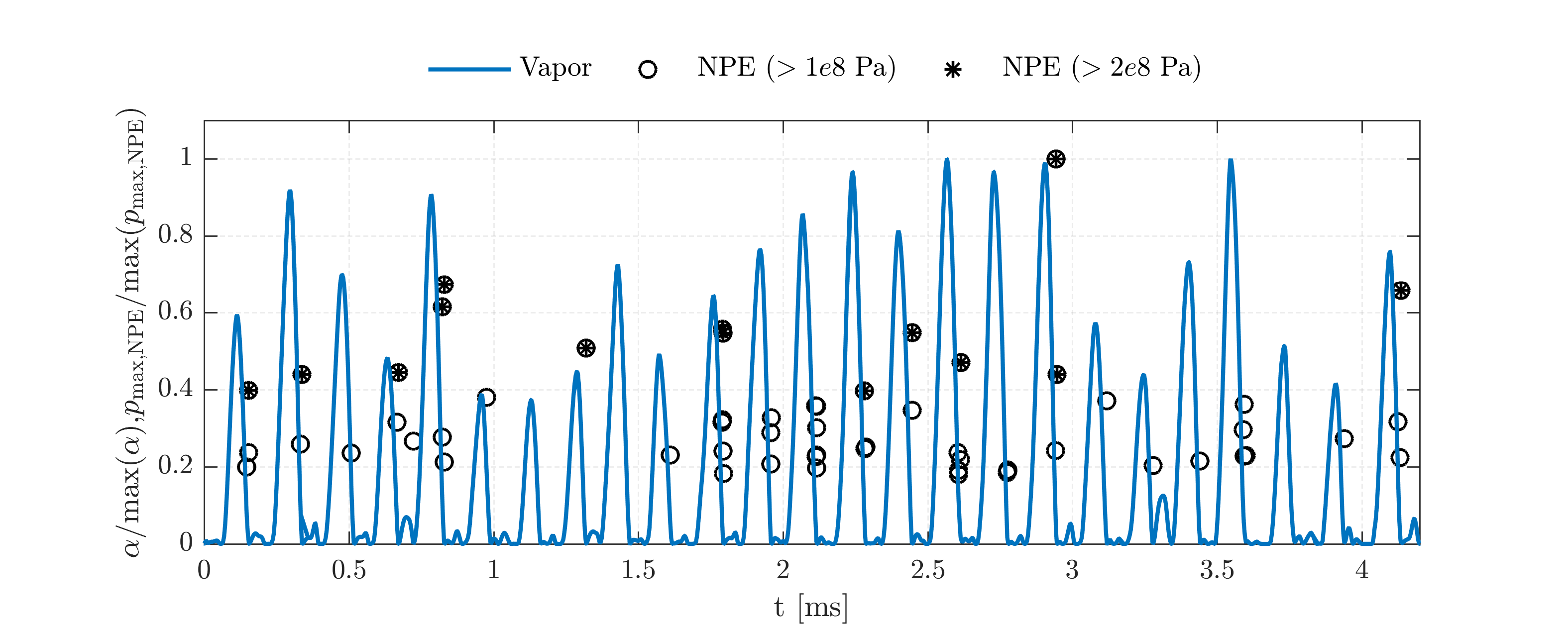}
      \caption{Vapor dynamics on the target and NPEs for \textit{OP}\,120\,bar. \new{Maximum pressure  and vapor content are normalized with the maximum values of the considered time interval.}}
       \label{fig:corr_vap_co_ev}
     \end{figure}

   \subsubsection{Temporal correlation}
     Surface loads on the target are correlated with collapsing vapor clouds, see also \cref{fig:ts_erosion}. \Cref{fig:corr_vap_co_ev} depicts the temporal evolution of the vapor content in the target region together with the NPEs. The plot underlines a clear temporal correlation of intensive surface loads with the cavitation dynamics, however not every cloud collapse directly results in the formation of an NPE. On average there is a potentially wall damaging event in about every second cycle, which is in \newnew{qualitative} agreement with experimental time-series observations of pit formation and cavitation cloud collapse by \citet{Fujisawa:2017jd}. 

  \FloatBarrier
  \pagebreak

    \begin{figure}[!tb]
      \centering
      \subfigure{\includegraphics[width=0.8\textwidth, trim={60 0 70 0},clip]{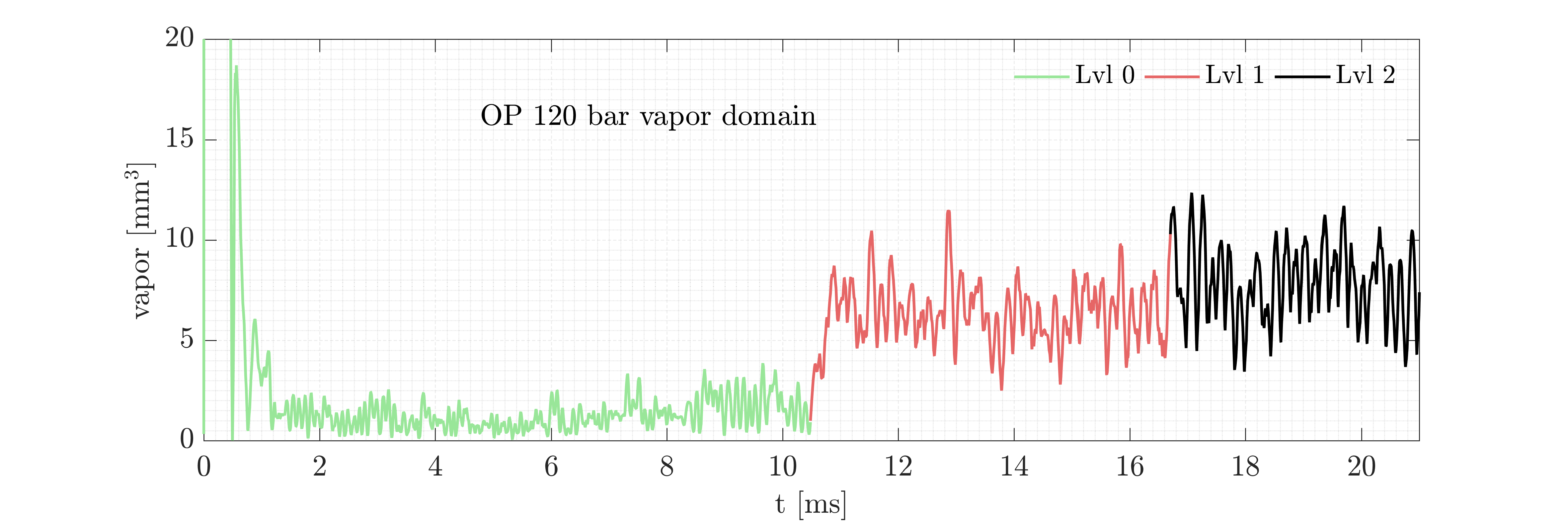}}
      \subfigure{\includegraphics[width=0.8\textwidth, trim={60 0 70 0},clip]{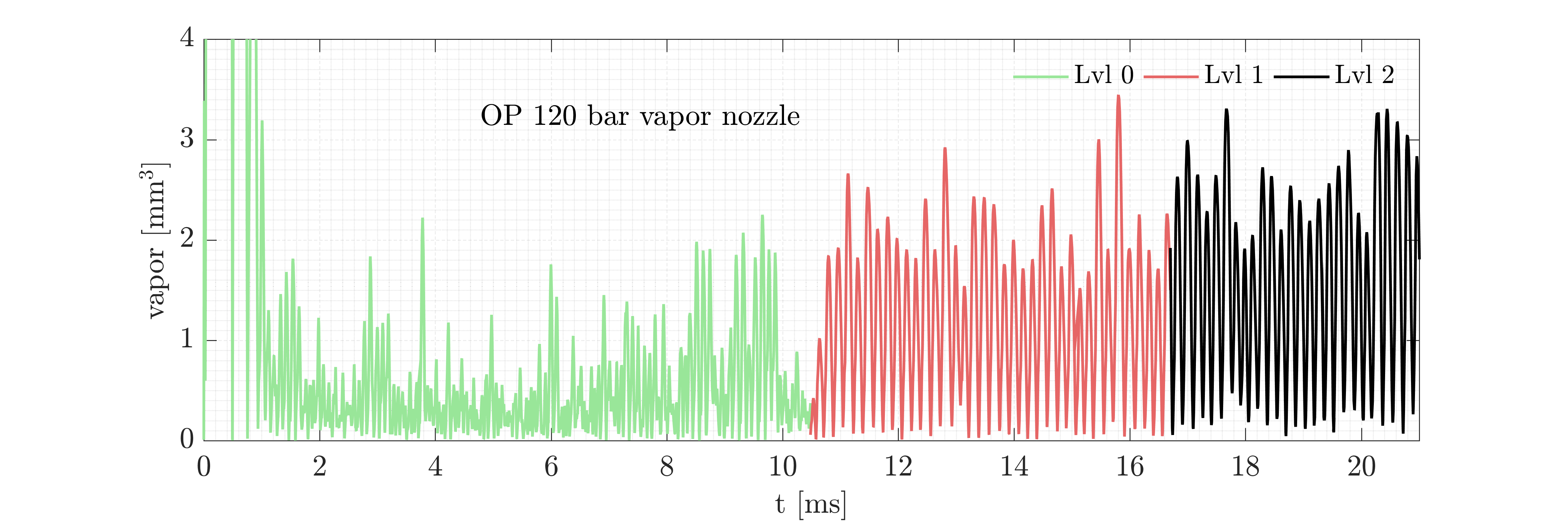}}
      \caption{Temporal evolution of the vapor content in the domain (top) and the diverging nozzle part (bottom) for different grid resolutions for \textit{OP}\,120\,bar.}
      \label{fig:grid_vapor}
    \end{figure}

    \section{Grid sensitivity study and scaling approach}\label{s:Grid} 

    We have conducted a grid sensitivity study by performing the simulations on three different grid levels, see \cref{tab:grid}. In this section, we evaluate the effect of the grid resolution and therefore compare the temporal evolution of the vapor content in the domain and the diverging nozzle, the distribution of the time-averaged vapor contents, and the predicted cavitation erosion aggressiveness. 

    The temporal development of the vapor content on different grid levels for \textit{OP}~120~bar is shown in \cref{fig:grid_vapor}, including the data for the divergent nozzle part and the entire domain. At the beginning of the simulations ($t=0$), there is a high vapor production, which then decreases as the flow field develops. With increasing grid resolution, the cavitation initiating processes are better captured and thus the vapor content increases, which has also been observed in previous studies (e.g. \citet{Orley:2015kt}). For both regions, we observe a convergence of the vapor content for \textit{Lvl}~1 and \textit{Lvl}~2. 

    \begin{figure}[!tb]
        \hspace{4cm}\subfigure[120 bar \textit{Lvl}~0]{\includegraphics[width=3cm, trim={320 220 300 100},clip]{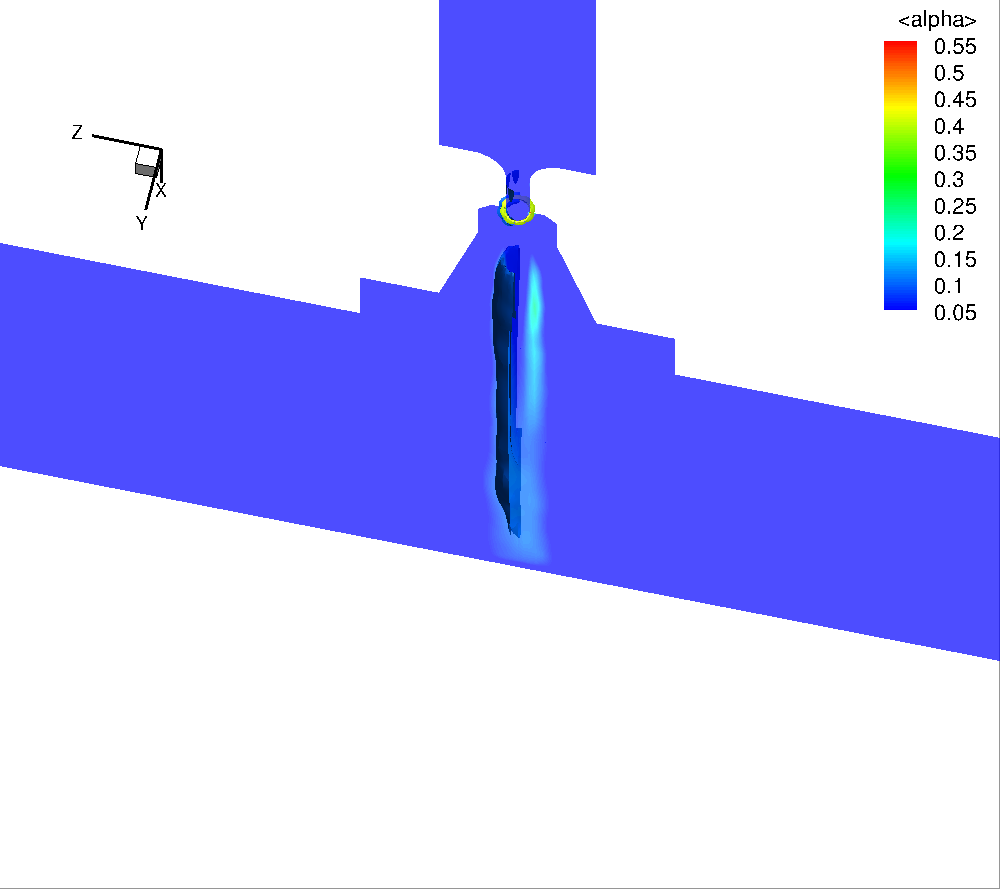}}
        \subfigure[\textit{Lvl}~1]{\includegraphics[width=3cm, trim={320 220 300 100},clip]{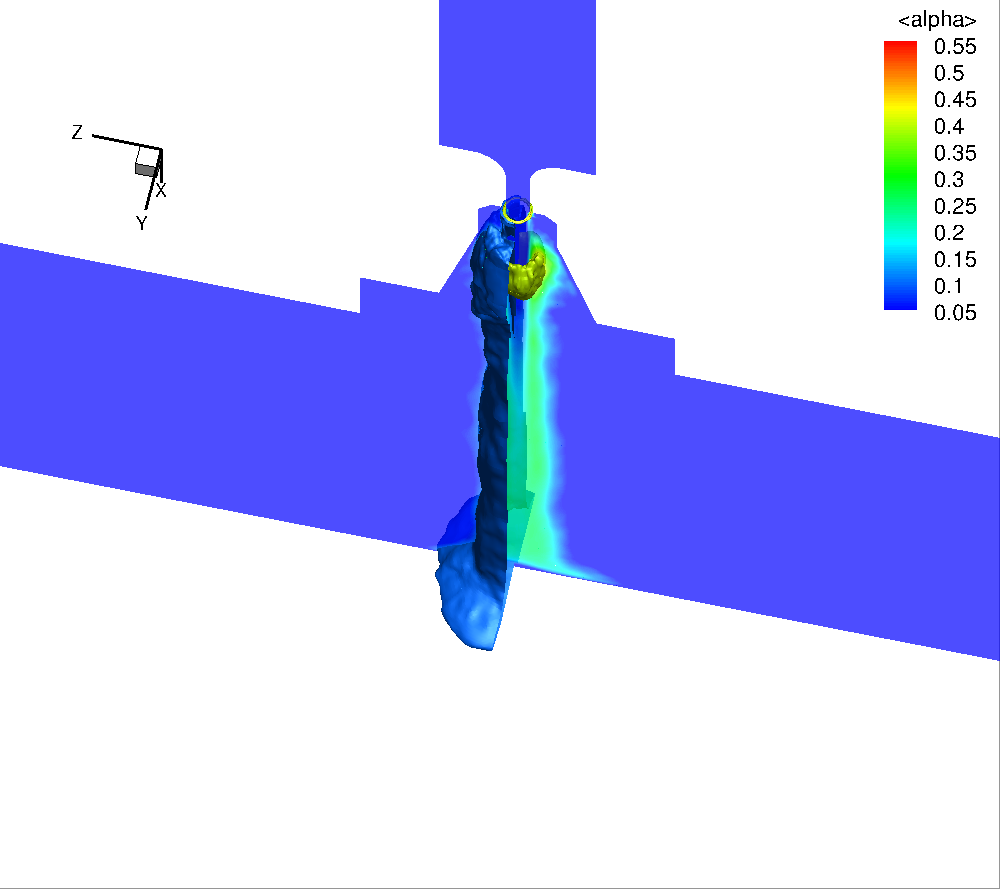}}
        \subfigure[\textit{Lvl}~2]{\includegraphics[width=3cm, trim={320 220 300 100},clip]{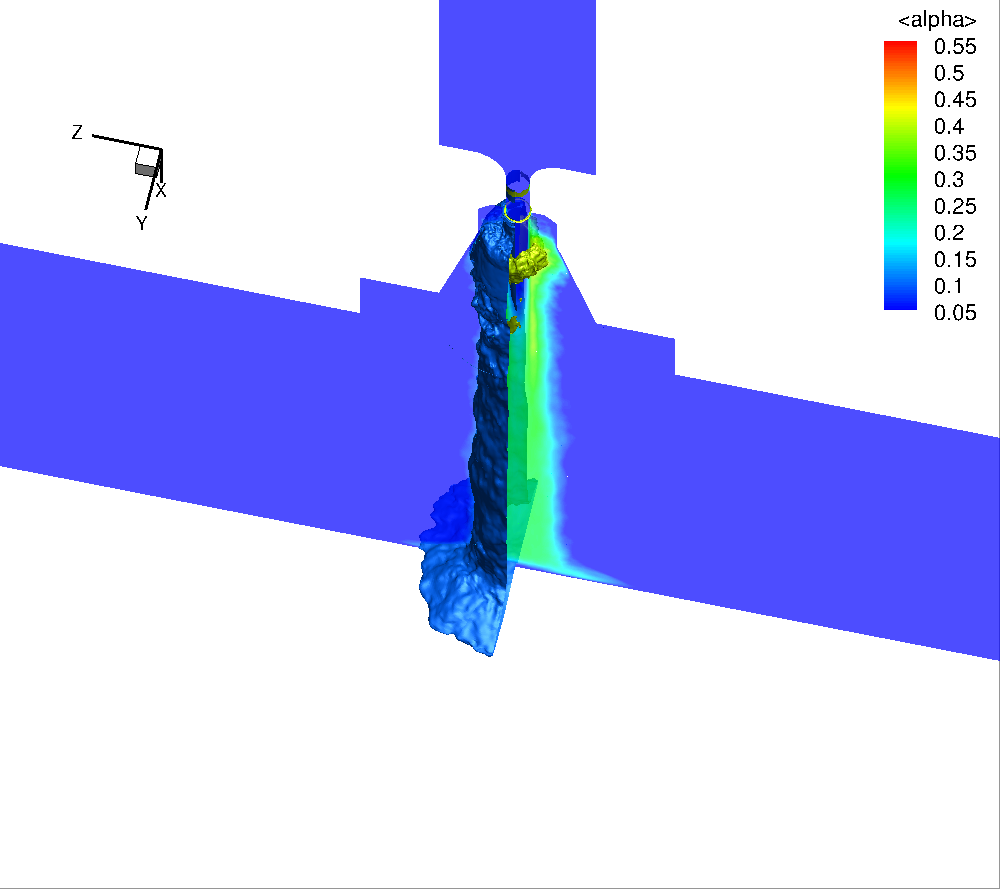}}
        \subfigure{\includegraphics[width=1.5cm]{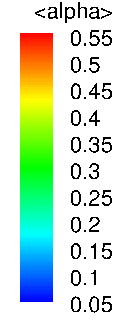}}\newline

        \hspace{4cm} \subfigure[201 bar \textit{Lvl}~0]{\includegraphics[width=3cm, trim={320 10 300 100},clip]{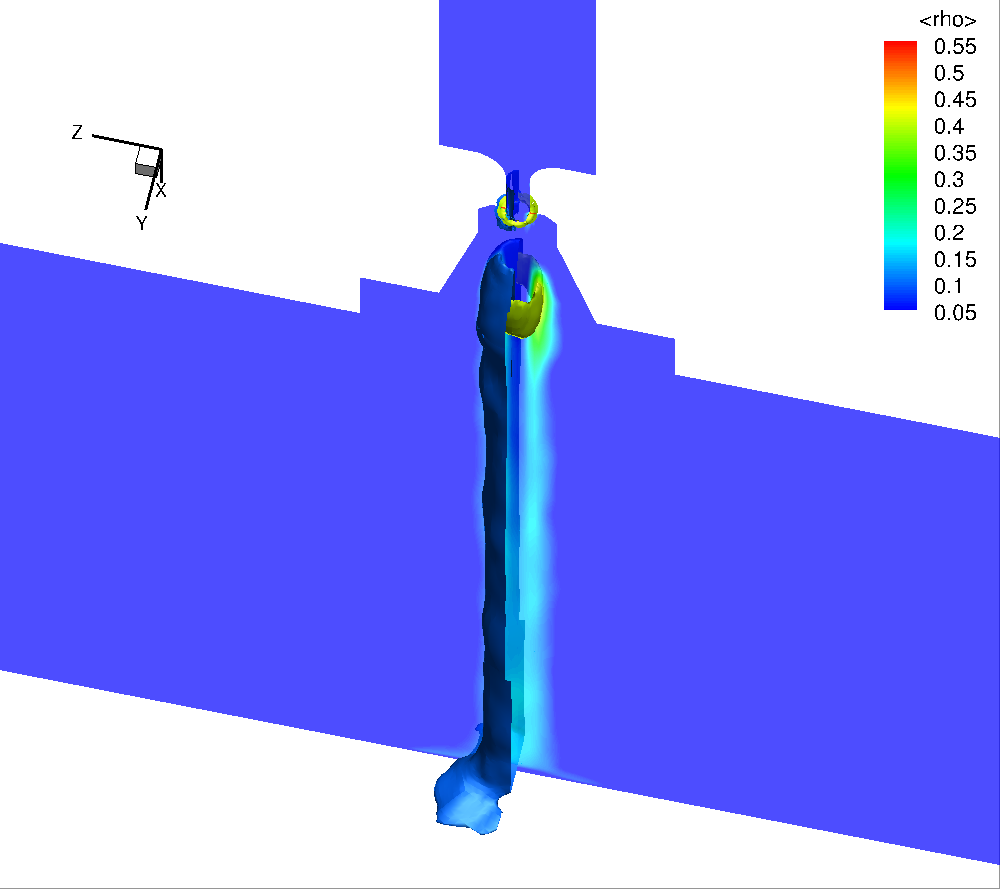}}
        \subfigure[\textit{Lvl}~1]{\includegraphics[width=3cm,trim={320 10 300 100},clip]{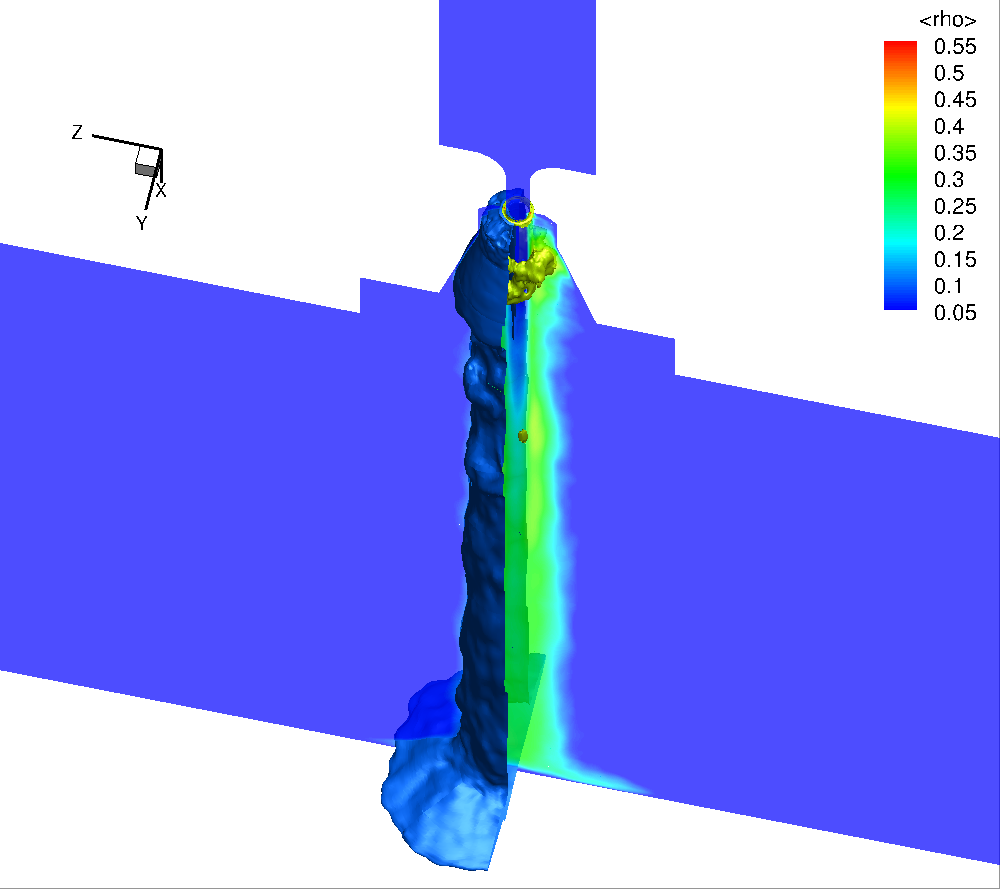}}
        \subfigure[\textit{Lvl}~2]{\includegraphics[width=3cm,trim={320 10 300 100},clip]{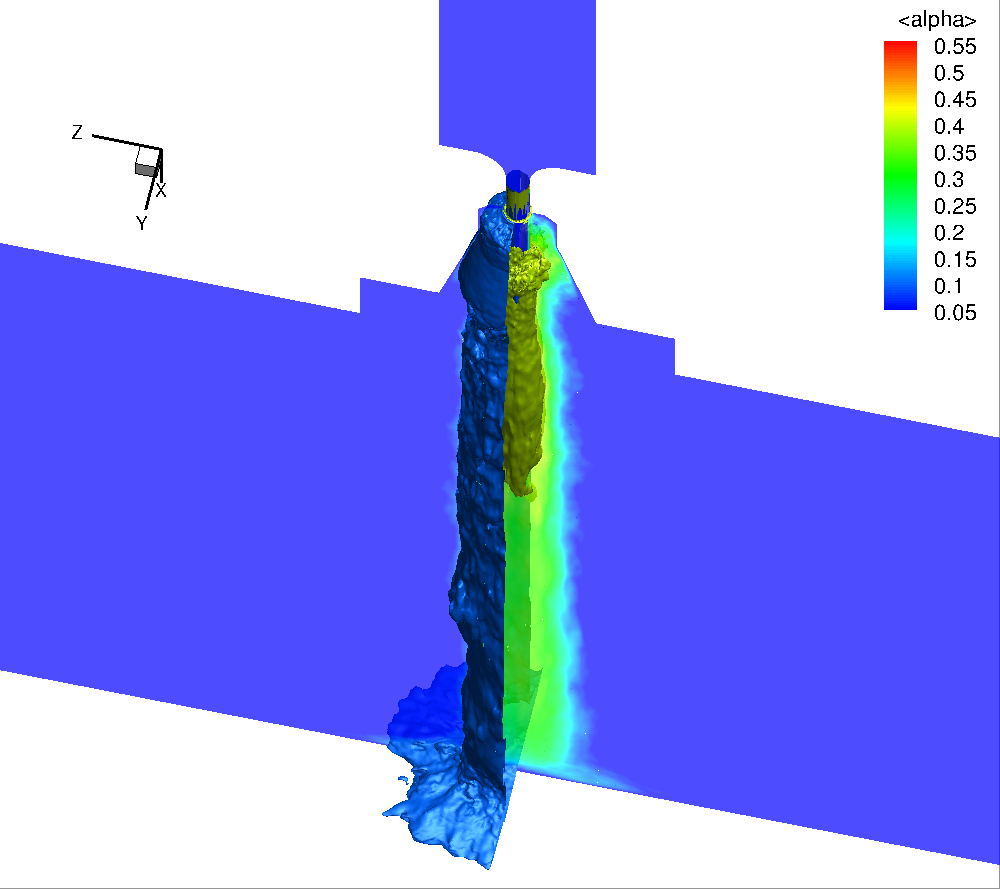}}
      \caption{Time-averaged vapor distribution for different grid resolutions. Top \textit{OP} 120 bar, bottom \textit{OP} 201 bar. Analyzing time of 4\,\si{ms}. Midplane together with isosurfaces vapor 10\% (\newnew{blue}) and 40\% (\newnew{yellow}). 
     }
      \label{fig:grid_study_vapor_jet}
    \end{figure} 

    \Cref{fig:grid_study_vapor_jet} visualizes the time-averaged vapor contents obtained by averaging over 4 ms. In agreement with the plots in \cref{fig:grid_vapor}, there is a significant difference in the vapor distribution between \textit{Lvl}~0 and \textit{Lvl}~1. This demonstrates even more clearly that a certain grid resolution is required to resolve cavitation processes. E.g., at \textit{OP} 120 bar, at the coarsest grid resolution, the cavitating jet does not even reach the target on average. Furthermore, the resolution affects the fragmentation of the vapor structures, see also \citet{Mihatsch:2015db,schmidt2014assessment}. In the time-averaged data (\cref{fig:grid_study_vapor_jet}) it can be seen that fragmentation and fluctuations increase with the grid resolution, and there is still an increase between \textit{Lvl}~1 and \textit{Lvl}~2. 

        \begin{figure}[!tb]
        \includegraphics[width=\linewidth]{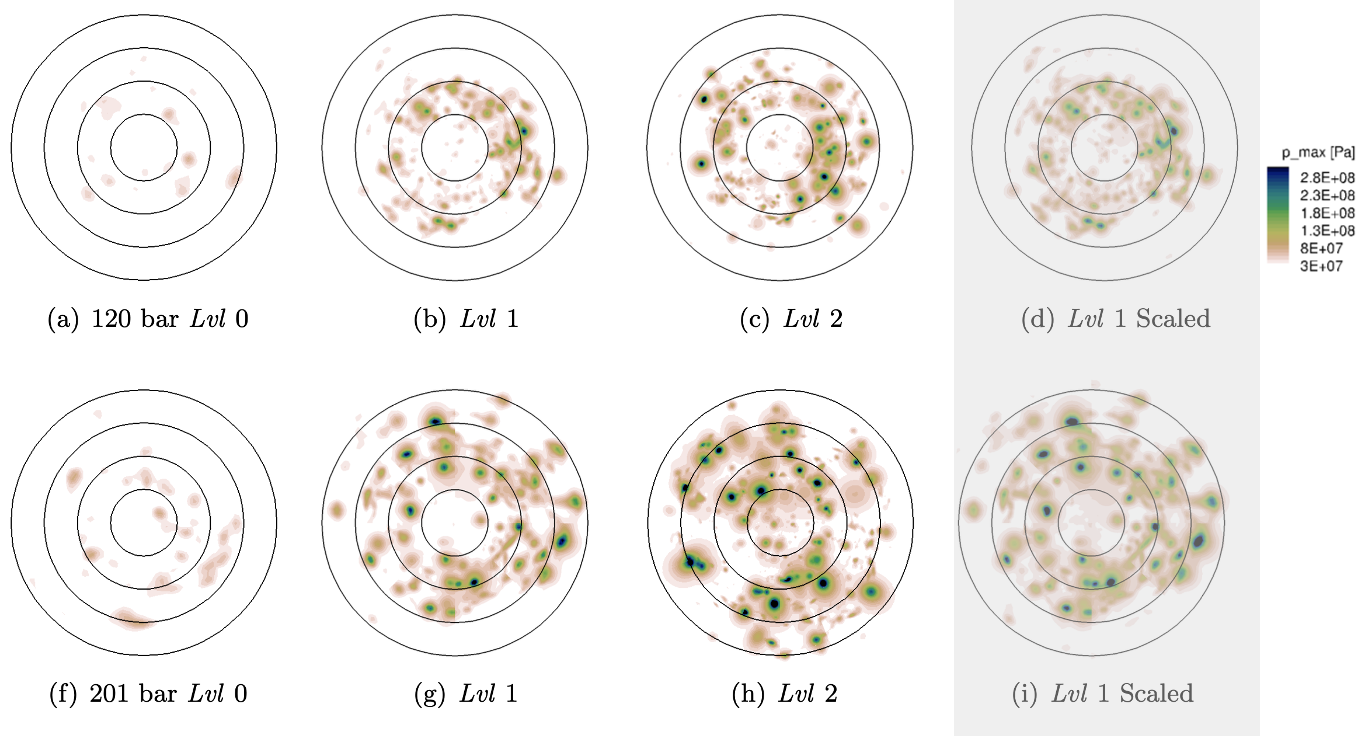}
      \caption{Recorded pressure peaks on the target for different grid resolutions and scaled maximum pressure distribution in the fourth column (\cref{eq:p_scale}). Top \textit{OP} 120 bar, bottom \textit{OP} 201 bar. Analyzing time of 4\,\si{ms}. Orientation rings at $r=1,2,3,4\,\si{mm}$. }
      \label{fig:grid_pmax}
    \end{figure} 

    \begin{table}[!tb]
    \caption{Comparison of pressure maxima on different grid resolutions. $\boldsymbol{\bar{p}_\mathrm{max,NPE}}$ refers to the average maximum pressure of the ten strongest NPEs. }
    \centering\begin{tabular}{|ccc|c|}
    \hline
      \textbf{Grid}&  
      \textbf{{\textit{OP}\,120\,bar} } $\boldsymbol{\bar{p}_\mathrm{max,NPE}}$\textbf{[Pa]} &
      \textbf{{\textit{OP}\,201\,bar} } $\boldsymbol{\bar{p}_\mathrm{max,NPE}}$\textbf{[Pa]} &
       \textbf{\textit{OP 201/OP 120}}\\
      \hline
      \textit{Lvl}~1 & 2.29e8  & 3.76e8 &  1.64   \\ 
      \textit{Lvl}~2 & 3.28e8  & 5.33e8 &  1.63   \\
      \hline 
      \textit{Lvl}~2 /\textit{Lvl}~1& 1.43  & 1.42 &    \\ 
      \hline
      \end{tabular}%
    \label{tab:grid_pmax}%
  \end{table}

  The grid resolution also affects the intensity of the recorded pressure peaks at collapse \citep{Mihatsch:2015db,Mihatsch:2017diss,schmidt2014assessment,Trummler:2021if}. The distribution of the maximum pressure on the wall is compared in \cref{fig:grid_pmax} for the different grid levels. Since both fragmentation and collapse-induced pressure peaks are grid-dependent, different maximum pressure distributions are obtained for the different grid resolutions. However, although the number and intensity of pressure peaks recorded are grid-dependent, the results allow for a qualitative prediction of the areas potentially prone to erosion damage.

\begin{figure}[!tb]
    \centering
       \new{\subfigure[]{\includegraphics[height=4cm, trim={0 0 0 0},clip]{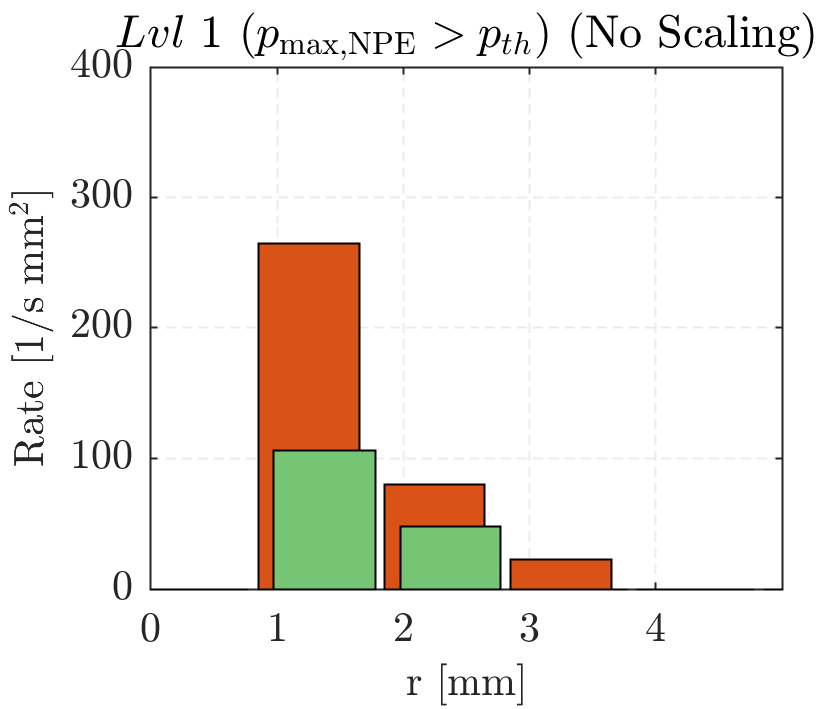}} }
       \subfigure[]{\includegraphics[height=4cm, trim={0 0 0 0},clip]{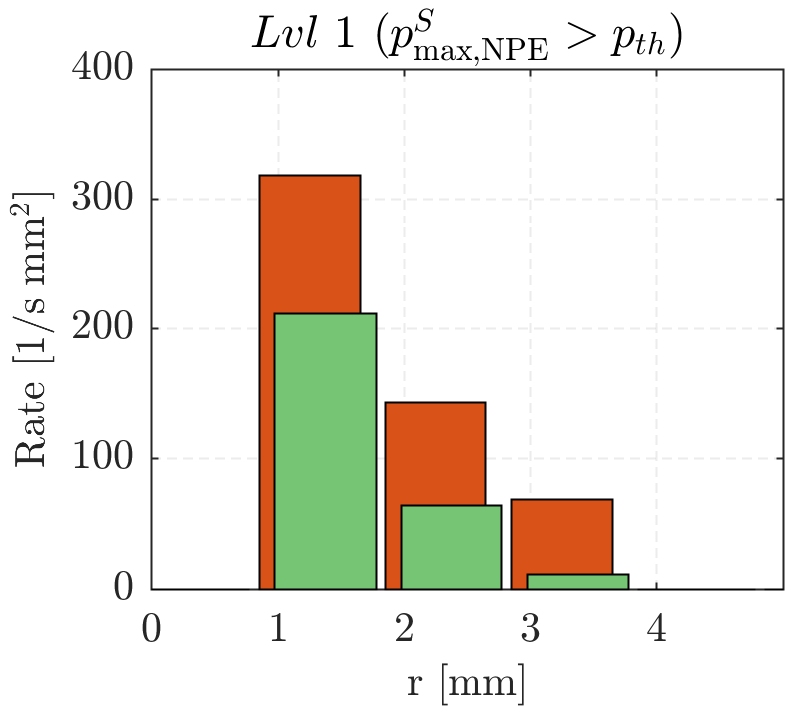}} 
       \subfigure[]{\includegraphics[height=4cm, trim={0 0 0 0},clip]{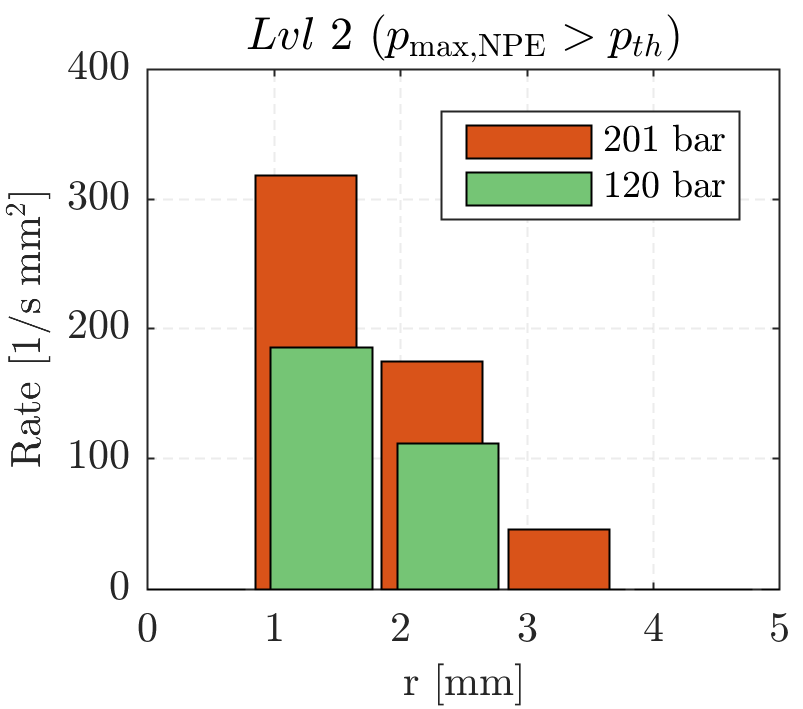}} 
      \caption{Pitting rates \new{(a) grid \textit{Lvl}~1 and no additional scaling, (b) \textit{Lvl}~1 and scaling using \cref{eq:p_scale}, (c) \textit{Lvl}~2}\newnew{.} }
      \label{fig:grid_pitting}
  \end{figure}

  To overcome grid dependence of the pressure impacts and to get comparable results from different grid resolutions, we propose a scaling method. We consider NPEs, since these are clearly defined and easy to handle. Following \citet{Mihatsch:2015db}, we assume a scaling in the form of 
        \begin{equation}
            p^S_\mathrm{max,NPE}=p_\mathrm{max,NPE} (l/l_\mathrm{0})^{\kappa}
            \label{eq:p_scale}
        \end{equation}
  where $l$ denotes the characteristic grid length, $l_\mathrm{0}$ a reference length, and $\kappa$ the scaling factor. For the evaluation shown here, we take the grid length of \textit{Lvl}~2 as reference length. To determine $\kappa$, we average the maximum pressure of the ten strongest NPE and compared them for different grid resolutions. The values are listed in \cref{tab:grid_pmax}. For both operating conditions, the ratio of the averaged maximum wall pressure on grid \textit{Lvl}~2 to grid \textit{Lvl}~1 is about 1.4. Thus, we approximate $\kappa=0.5$ which scales the pressure peaks recorded on Grid \textit{Lvl}~1 by a factor of 1.41 ($\left(l_{Lvl~1}/l_{Lvl~2}\right)^{\kappa}=2^{\,0.5}$). The scaled maximum wall pressure distribution of \textit{Lvl}~1 is depicted in the fourth column in \cref{fig:grid_pmax}. \new{Due to scaling, NPEs on \textit{Lvl}~1 are more pronounced and their spatial extent more closely resembles that of NPEs on \textit{Lvl}~2.} \Cref{fig:grid_pitting} compares the pitting rate obtained on \textit{Lvl}~1 \new{with no scaling (\cref{fig:grid_pitting}~(a)) and on \textit{Lvl}~1 using the proposed scaling (\cref{fig:grid_pitting}~(b))} with that on \textit{Lvl}~2 \new{(\cref{fig:grid_pitting}~(c)). Using the proposed scaling, the scaled pitting rates obtained on \textit{Lvl}~1 are significantly closer to those on \textit{Lvl}~2, see \cref{fig:grid_pitting}~(b,c).} Conveniently, the scaling can also be applied a-posteriori since recorded events can simply be filtered by a certain threshold value. For more detailed analyses, like the ones presented in \cref{ss:Erosion_prediction}, more caution has to be taken since the presented scaling also leads to an increase of the relative NPE size. 

  Additionally, it is worth noting that the ratio of the highest pressure peaks at {\textit{OP}\,201\,bar} to those at {\textit{OP}\,120\,bar} is on both grid resolutions about the same, see \cref{tab:grid_pmax}. This further underlines that simulation results obtained on identical grid resolutions allow for a comparison of the erosion intensity of different operating conditions. 

\section{Conclusions}\label{s:Conclusion}

  This paper focuses on numerical erosion prediction. We have performed fully compressible, high-resolution numerical simulations of a cavitating liquid jet featuring two different operating points. 

  \new{
    For the numerical erosion prediction, we have introduced a new evaluation method termed \textit{numerical pit equivalents} (NPEs). These are clustered cell-based pressure maxima designed to represent a numerical counterpart to experimental pits. Using the NPE data, we have demonstrated the versatile evaluation methods therewith, such as the evaluation of pitting rates, and gained new insights into the duration of pit formation. 

    The main benefit of NPEs lies in the additional time information. This can be particularly important for understanding erosion mechanisms in large configurations that are difficult to access experimentally, such as ship propellers. Here, numerical simulations using NPEs could provide more detailed insight into the damage mechanisms, for example, by revealing if there are a few strong pressure loads or a superposition of many loads. Overall, the generation of NPEs based on hydrodynamic pressure data can be seen as another step towards quantitative numerical erosion prediction. Nevertheless, for the estimation of actual cavitation erosion, the material response to the pressure loads has to be incorporated. This has been successfully achieved for near-wall bubble collapses by e.g. \citet{hsiao2014modelling,sarkar2021fluid} using coupled fluid-structure-interaction simulations. For NPEs, augmenting the data output with relevant material parameters could be a way to determine material response a-posteriori, similar to the approach of e.g. \citet{kaufhold2017numerical}.

    Additionally, we have also presented a comprehensive grid study using three different grid resolutions. This study revealed that especially the maximum wall pressure exhibits a strong grid dependence. To overcome this dependence, we proposed a simple scaling method that leads to comparable maximum pressure distributions and pitting rates of NPEs on different grid resolutions. To take another step towards grid-independent prediction, we plan to extend our approach using the erosion risk indicators suggested by \citet{Brunhart:2019ge}. 
  }

\section*{Declaration of Competing Interest}

  The authors declare that they have no known competing financial interests or personal relationships that could have appeared to influence the work reported in this paper.

\section*{Acknowledgment}

  We thank the Robert Bosch GmbH for providing the geometrical data of the considered setup and the thermodynamic properties of the test fluid. 
  Further, we gratefully acknowledge the Gauss Centre for Supercomputing e.V. (www.gauss-centre.eu) for funding this project by providing computing time on the GCS Supercomputers SuperMUC and SuperMUC-NG at Leibniz Supercomputing Centre (www.lrz.de). 

\FloatBarrier

\section*{References}
  \bibliography{main}
  \bibliographystyle{elsarticle-harv}
  \biboptions{authoryear}

\end{document}